\documentclass[11pt]{JHEP3}

\usepackage{amsmath}
\usepackage{amssymb}
\usepackage{graphicx}
\usepackage{cite}

\newcommand{\be}{\begin{eqnarray}}
\newcommand{\ee}{\end{eqnarray}}
\newcommand{\nn}{\nonumber}
\newcommand{\bn}{\begin{enumerate}}
\newcommand{\en}{\end{enumerate}}

\def\Ds{D \hskip -7pt / \hskip 2pt}

\def\IC{\mathbb{C}}

\def\IZ{\mathbb{Z}}

\def\CJ{{\cal J}}

\def\CM{{\cal M}}
\def\CN{{\cal N}}

\def\CR{{\cal R}}

\def\a{\alpha}
\def\b{\beta}
\def\g{\gamma}
\def\d{\delta}
\def\e{\epsilon}

\def\l{\lambda}
\def\m{\mu}
\def\n{\nu}

\def\r{\rho}

\def\s{\sigma}

\def\t{\tau}

\def\w{\omega}
\def\dd{\rm d}

\def\half{\frac{1}{2}}

\def\Tr{{\rm Tr}}

\def\da{{\dot{\a}}}
\def\db{{\dot{\b}}}
\def\dg{{\dot{\g}}}
\def\dd{{\dot{\d}}}

\def\tps{{\tilde{\psi}}}

\newcommand{\sla}[1]{{/\!\!\!\!{#1}}}

\newcommand{\ZZ}{{\mathbb Z}}
\newcommand{\CC}{{\mathbb C}}
\newcommand{\RR}{{\mathbb R}}

\def\dda{{\dot{a}}}
\def\ddb{{\dot{b}}}
\def\ddc{{\dot{c}}}
\def\ddd{{\dot{d}}}

\def\Ua{\underline{a}}
\def\Ub{\underline{b}}
\def\Uc{\underline{c}}
\def\Ud{\underline{d}}

\def\Uda{\underline{\dot{a}}}
\def\Udb{\underline{\dot{b}}}
\def\Udc{\underline{\dot{c}}}
\def\Udd{\underline{\dot{d}}}

\title{ $\CN=5,6$ Superconformal Chern-Simons Theories \\
 and M2-branes on Orbifolds}

\author{Kazuo Hosomichi$^1$, Ki-Myeong Lee$^1$, Sangmin Lee$^2$, Sungjay Lee$^1$ and Jaemo Park$^{3,4,5}$

\\
\\
$^1$Korea Institute for Advanced Study, Seoul 130-012, Korea
\\
$^2$Department of Physics \& Astronomy, Seoul National University,
Seoul 151-747, Korea
\\
$^3$Department of Physics, POSTECH,
Pohang 790-784, Korea
\\
$^4$Postech Center for Theoretical Physics (PCTP), Postech, Pohang
  790-784, Korea
\\$^5$Department of Physics, Stanford University, Stanford, CA
94305-4060, USA
\\
\\
E-mail:
\email{hosomiti@kias.re.kr, klee@kias.re.kr, sangmin@snu.ac.kr, sjlee@kias.re.kr, jaemo@postech.ac.kr}
}

\abstract{We explore further our recent generalization of the
$\CN=4$ superconformal Chern-Simons theories of Gaiotto and Witten. 
We find and construct explicitly theories of enhanced
$\CN=5$ or $6$ supersymmetry, especially $\CN=5$, $Sp(2M)\times
O(N)$ and $\CN=6$, $Sp(2M)\times O(2)$ theories. The
$U(M)\times U(N)$ theory coincides with the $\CN=6$ theory
of Aharony, Bergman, Jafferis and Maldacena (ABJM).  We argue that
the $\CN=5$ theory with $Sp(2N)\times O(2N)$ gauge group can be
understood as an orientifolding of the ABJM model with $U(2N)\times
U(2N)$ gauge group. We briefly discuss the Type IIB brane
construction of the $\CN=5$ theory and the geometry of the
M-theory orbifold. }

\preprint{KIAS-P08046 \\ SU-ITP-08/13}

\begin{document}

\section{Introduction and Concluding Remarks}

Superconformal Chern-Simons theories have become a subject of
intensive research recently. Schwarz \cite{sch} suggested that
Chern-Simons theories without Yang-Mills kinetic term may be used
to describe the $\CN=8$ superconformal M2-brane world-volume
theory. The idea was crystalized by Bagger and Lambert
\cite{BL1,BL2,BL3} and Gustavsson \cite{gus1,gus2} (BLG) who
proposed an $\CN=8$  Chern-Simons theory based on   3-algebra, and
gave an explicit example with $SO(4)=SU(2)\times SU(2)$ gauge
group. The $SO(4)$ BLG theory can be rewritten as an ordinary
gauge theory \cite{sc2,rams}. It is conjectured to be a specific
theory of some M2 brane configuration~\cite{tong,mukhi}.

More recently, Aharony, Bergman, Jafferis and Maldacena (ABJM)
\cite{ABJM} have found $\CN=6$ superconformal Chern-Simons
theories of $U(M)\times U(N)$ gauge group, and have argued that
the theories with $U(N)\times U(N)$ gauge group and Chern-Simons
level $k$ is the holographic dual of the M-theory geometry
background arising from $N$ M2 branes on the orbifold
$\IC^4/\IZ_k$. Especially for $k=1,2$ cases, it has been argued
that the supersymmetry is enhanced to $\CN=8$.

On the other hand, with somewhat different motivation, Gaiotto and
Witten \cite{gw} have constructed a class of $\CN=4$
superconformal Chern-Simons theories coupled to hyper-multiplets.
The gauge group and matter content  are severally restricted and
determined with the help of general classification of Lie
super-algebra. In particular, the theory can come with $U(M)\times
U(N)$ or $Sp(2M)\times O(N)$ gauge group and a single
bi-fundamental hyper-multiplet. In a subsequent work \cite{hl3p},
we have constructed more general $\CN=4$ linear quiver-type
theories where bi-fundamental hyper and twisted multiplets
alternate between connected nodes. The special case of
$SU(2)\times SU(2)$ gauge group with both   hyper and   twisted
hyper-multiplets becomes identical to the BLG theory with enhanced
$\CN=8$ supersymmetry. It was also suggested in \cite{hl3p} that
the embedding of the Chern-Simons theories into IIB string theory
\cite{gw} can be T-dualized to make contact with M2-branes on
orbifolds.

In this work, we show by an explicit construction that  the
$\CN=4$ theories with two gauge groups and both hyper and twisted
hyper-multiplets in the same gauge representation always have an
enhanced $\CN=5$ or more supersymmetry. Especially those with
$U(M)\times U(N)$ gauge group coincides with the $\CN=6$ ABJM
theory. We also find new $\CN=5$ theories of $Sp(2M)\times O(N)$
($N>2$) and $\CN=6$ theories of $Sp(2M)\times O(2)$. We argue that
the $Sp(2N)\times O(2N)$ theory can be  obtained by a simple
orientifolding of $\CN=6 $ $U(2N)\times U(2N)$ ABJM theory and can
be regarded as the holographic dual of the M-theory geometry for
M2 branes exploring the orbifold $\IC^4/D_k$ where $2k-2$ is the
level of the Chern-Simons coefficient.

The $\CN=3$ theories can come with arbitrary gauge group and
matter hyper-multiplets, and are not subject to any
quantum correction to  the Chern-Simons level. The superconformal
theory of $\CN=2,3$ theories has been studied extensively by
Gaiotto and Yin~\cite{gy}. ABJM have shown that the $\CN=3$ theory
with $U(M)\times U(N)$ gauge group and a `pair' of bi-fundamental
matter field have an enhanced $\CN=6$ supersymmetry, by arguing
that the theory has a $SU(2)\times SU(2)$ global symmetry which
does not commute with the  $SU(2)$ $R$-symmetry. The ABJM theory
falls into our category of theories with enhanced supersymmetry.

We begin with a brief summary of $\CN=4$ theories with both types
of hyper-multiplets in Sec.~\ref{n45}, and then show by an
explicit construction that whenever the hyper-multiplets belong to
the same gauge representation, there is an enhancement of the
supersymmetry to at least $\CN=5$. Then we work out the Lagrangian
for the theories of  $Sp(2M)\times O(N)$ gauge group in detail.

In Sec.~\ref{n6}, we study a further enhancement of $\CN=5$ to
$\CN=6$ supersymmetry, which occurs whenever the matter
representation can be decomposed a complex representation $(R)$
and its complex conjugate representation $(\bar{R})$, or the
matter representation is purely real. The $U(M)\times U(N)$ gauge
theory becomes the ABJM model, and the $Sp(2M)\times O(2)$ gauge
theory provides a new example of $\CN=6$ superconformal field
theories.

We could pursue our analysis further and construct general
$\CN=7,8$ theories as well. {}From the $\CN=6$ point of view,
enhancement of supersymmetry to $\CN \ge 7$ requires that the
representation $R$ be real; the $SO(4)$ BLG theory is such an
example. Hopefully, there could be other cases where the matter
representation is real besides the BLG case.

Given the IIB brane configuration of the ABJM model, it is easy to
relate   our new $\CN=5$ $OSp(M|N)$ theories   to an orientifold
of the ABJM model. We elaborate on this point in Sec.~\ref{ori}.
Taking the T-duality to M-theory as in the ABJM model, we obtain a
$D_k$ orbifold $\IC^4$. Orbifolds of $\IC^4$ preserving $\CN=5$
supersymmetry were discussed earlier in \cite{a0} and very
recently in \cite{hana}.

In Appendices, we provide the detailed computations for the
$\CN=5,6$ cases and express the mass deformation of the $\CN=4$
Lagrangian~\cite{hl3p} in $\CN=5,6$ language. It shows that 
all of the $\CN=5,6$ supersymmetries are preserved, 
while the $SO(5)$ or $SO(6)$ $R$-symmetries are partially broken to
$SO(4)$ or $SO(4)\times U(1)$, respectively. This is somewhat
expected from the mass deformation of the BLG
theory~\cite{x3,hll}.

We close this introduction with some directions for further study.
The 3-algebra structure played a crucial role in making the BLG
model compatible with $\CN=8$ supersymmetry, while the ABJM model
at $k=1$ is argued to have $\CN=8$ without using the 3-algebra
structure. It would be desirable to understand the relation
between the two approaches. A recent paper \cite{kek} has taken a
step in this direction (see also
\cite{bf1,bf2,bf3,bx1,bb1,bx2,bb2}). More work is required to
establish the AdS/CFT correspondence of the Chern-Simons theories.
The relevant topics include extension to $\CN=4$ orbifolds
\cite{a2,ima}, superconformal indices \cite{a8},
Penrose limit \cite{a10},  integrability \cite{mz}, non-supersymmetric generalization \cite{adi}, and partition function
\cite{hana}. Finally, it remains to derive, from first principles,
much richer family of $\CN\ge 2$ superconformal theories with
known M-theory geometry \cite{a1,a2,mc1,mc2,mc3,lyee}.

\section{$\CN=5$ Superconformal Theories}
\label{n45}

\subsection{Review of $\mathbf{\CN=4}$}
\label{n4}

We review the construction of general $\CN=4$ superconformal
Chern-Simons-matter theories \cite{gw,hl3p}. We start with an
$Sp(2n)$ group and let $A,B$ indices run over a $2n$-dimensional
representation. We denote the anti-symmetric invariant tensor of
$Sp(2n)$ by $\omega_{AB}$ and choose all the generators $t^A_{~B}$
to be anti-Hermitian $(2n\times 2n)$ matrices  such that
$t_{AB}=\omega_{AC}t^C_{~B}$ are symmetric matrices. We consider a
Chern-Simons gauge theory whose gauge group is a subgroup of
$Sp(2n)$ and   denote  anti-Hermitian generators of the gauge
group as $(t^m)^A_{~B}$ which satisfy
\[
 [t^m,t^n]=f^{mn}{}_p t^p.
\]
The gauge field is denoted by $(A_m)_\mu$, and the
adjoint indices are raised or lowered by an invariant quadratic
form $k^{mn}$ or its inverse $k_{mn}$ of the gauge group.

We couple the gauge theory with hyper and twisted
hyper-multiplet matter fields
 $(q^A_\a, \psi^A_\da ; \tilde{q}^A_\da,\tilde{\psi}^A_\a)$
 satisfying the reality conditions
\be
 \bar q_A^\a=(q^A_\a)^\dagger=\epsilon^{\a\b}\omega_{AB}q^B_\b,~~~~~~
 \bar \psi_\da^A=(\psi^A_\da)^\dagger=\epsilon^{\da\db}\omega_{AB}\psi^B_\db,
\label{real4} \ee
and similar conditions for $\tilde{q}_\da^A$ and
$\tilde{\psi}_\a^A$. We use $(\a,\b ; \da,\db)$ doublet indices
for the $SU(2)_L\times SU(2)_R$ $R$-symmetry group. We also have
the inverse tensors, $\omega^{AB},\epsilon_{\alpha\beta},
\epsilon_{\dot{\a}\dot{\b}}$ such that, say, $\omega_{AC}
\omega^{CB}=\delta_A^{\ B}$, and $\epsilon^{\a\g}\epsilon_{\g\b}=
\delta^\a_{\ \b}$.

Both types of hyper-multiplets share the same gauge symmetry, so
the structure constants $f^{mn}_{~~~p}$ and the quadratic form
$k^{mn}$ are identical. But, they can take different
representations   in general $\CN=4$ theories. For $\CN>4$
supersymmetry, however, the two types of hyper-multiplets  should
be combined together into a bigger multiplet, so they have to take
the same representation.

In the construction of Ref.~\cite{gw,hl3p}, $\CN=1$ super-field
formulation was used and conditions on the super-potentials for
enhancement to $\CN=4$ was examined. It was found that there is
essentially one constraint equation (called ``fundamental
identity'' in \cite{gw}),
\be k_{mn} t^m_{A(B} t^n_{CD)} \; =\; 0,
\ee
where the expression is summed over the cyclic permutation of
indices $B,C,D$. When this condition is satisfied, all $\CN=1$
super-potentials are uniquely determined, and we end up with an
$\CN=4$ theory.

Following \cite{gw}, we introduce the ``moment map'' and ``current'' operators,
\be
\m^m_{\a\b} \equiv t^m_{AB} q^A_\a q^B_\b
, ~~~~
\jmath^m_{\a \da} \equiv q^A_\a t^m_{AB} \psi^B_\da
, ~~~~
\tilde{\m}^m_{\da\db} \equiv \tilde{t}^m_{AB} \tilde{q}^A_\da \tilde{q}^B_\db
, ~~~~
\tilde{\jmath}^m_{\da \a} \equiv \tilde{q}^A_\da \tilde{t}^m_{AB} \tilde{\psi}^B_\a .
\ee
Using these notations, we can write down
the Lagrangian of general $\CN=4$ superconformal
Chern-Simons-matter theories in a fully covariant form \cite{hl3p},
\begin{eqnarray}
 {\cal L} &=&
 \frac{\varepsilon^{\m\n\l}}{4\pi}
 \left(k_{mn} A^m_\mu \partial_\n A^n_\l
      +\frac13 f_{mnp} A^m_\m A^n_\n A^p_\l \right)
\nn\\&&
 +\frac12\omega_{AB}
  \left(-\epsilon^{\a\b} D q^A_\a D q^B_\b
        +i\epsilon^{\da\db}\psi^A_\da \Ds\psi^B_\db \right)
 +\frac12 \omega_{AB}
  \left(-\epsilon^{\da\db} D \tilde q^A_\da D \tilde q^B_\db
       +i\epsilon^{\a\b}\tilde\psi^A_\a \Ds\tilde\psi^B_\b \right)
 \nn\\&&
 -i\pi k_{mn} \e^{\a\b} \e^{\dg\dd}\jmath^m_{\a\dg}\jmath^n_{\b\dd}
 -i\pi k_{mn} \e^{\da\db} \e^{\g\d}\tilde\jmath^m_{\da\g}\tilde\jmath^n_{\db\d}
 +4\pi ik_{mn}\epsilon^{\a\g}\epsilon^{\db\dd}
   \jmath^m_{\a\db}\tilde\jmath^n_{\dd\g}
 \nn\\&&
 +i\pi k_{mn}\left(
 \epsilon^{\da\dg}\epsilon^{\db\dd}
 \tilde\mu^m_{\da\db} \psi_\dg^At^n_{AB}\psi_\dd^B
 +\epsilon^{\a\g}\epsilon^{\b\d}
  \mu^m_{\a\b}\tps_\g^A\tilde t^n_{AB}\tps_\d^B
 \right)
 \nn\\&&
  -\frac{\pi^2}6f_{mnp}(\mu^m)^\a_{~\b}(\mu^n)^\b_{~\g}(\mu^p)^\g_{~\a}
  -\frac{\pi^2}6f_{mnp}
   (\tilde\mu^m)^{\dot{\a}}_{~\dot{\b}}
   (\tilde\mu^n)^{\dot{\b}}_{~\dot{\g}}
   (\tilde\mu^p)^{\dot{\g}}_{~\dot{\a}}
 \nn\\&&
  +\pi^2(\tilde\mu^{mn})^\dg_{~\dg}(\mu_m)^\a_{~\b}(\mu_n)^\b_{~\a}
  +\pi^2(\mu^{mn})^\g_{~\g}(\tilde\mu_m)^\da_{~\db}(\tilde\mu_n)^\db_{~\da}\,.
\label{Lful4}
\end{eqnarray}
The supersymmetry transformation law reads,
\begin{eqnarray}
 && \delta q_\alpha^A
    = +i\eta_\alpha^{~\;\dot\alpha}\psi_{\dot\alpha}^A,
~~~
    \delta\tilde q_{\dot\alpha}^A
    = -i\eta_{~\;\dot\alpha}^{ \alpha}\tilde\psi_\alpha^A,
~~~
    \delta A^m_\mu = 2\pi i\eta^{\alpha\dot\alpha}\gamma_\mu
            (\jmath^m_{\alpha\dot\alpha}-\tilde\jmath^m_{\dot\alpha\alpha}),
 \nn\\
 && \delta\psi_{\dot\alpha}^A
    = +\left[\sla Dq_\alpha^A
           +\frac{2\pi}3(t_m)^A_{~B}q^B_\beta(\mu^m)^\beta_{~\alpha}\right]
            \eta^\alpha_{~\;\dot\alpha}
           -2\pi(t_m)^A_{~B}q^B_\beta(\tilde\mu^m)^{\dot\beta}_{~\dot\alpha}
            \eta^\beta_{~\;\dot\beta} ,
 \nn\\
 && \delta\tilde\psi_\alpha^A
    = -\left[\sla D\tilde q_{\dot\alpha}^A
            +\frac{2\pi}3(\tilde t_m)^A_{~B}\tilde q^B_{\dot\beta}
             (\tilde\mu^m)^{\dot\beta}_{~\dot\alpha}\right]
            \eta_\alpha^{~\;\dot\alpha}
           +2\pi(\tilde t_m)^A_{~B}\tilde q^B_{\dot\beta}
             (\mu^m)^{\beta}_{~\alpha}
            \eta_\beta^{~\;\dot\beta} .
            \label{Sful4}
\end{eqnarray}
The spinor parameter $\eta_{\a\db}$ satisfies the reality
condition
\be (\eta_\a^{\ \dot{\b}})^*= \eta^\a_{\ \dot{\b}}
=\epsilon^{\a\b}\epsilon_{\db\da}\eta_\b^{\ \da} \, . \ee

In Ref.~\cite{gw}, it was noticed that the fundamental identity can be
understood as the Jacobi identity for three fermionic generators
of a Lie super-algebra,
\begin{equation}
  [M^m,M^n]=f^{mn}_{~~~p}M^p,~~~~
  [M^m,Q_A]=Q_B (t^m)^B_{~A},~~~~
  \{Q_A,Q_B\}= t^m_{AB} M_m \,.
\end{equation}
This turns out to be a rather strong constraint on the field
content of the theory. Namely, the gauge group and matter should
be such that the gauge symmetry algebra can be extended to a
Lie super-algebra by adding fermionic generators associated to
hyper-multiplets.

The notion of Lie super-algebra characterizing $\CN=4$ theories
will be useful throughout the rest of the paper,
as we investigate the conditions for
enhanced supersymmetry ($\CN>4$).

\subsection{General construction}
\label{n5}

A necessary condition for supersymmetry enhancement is that the two
types of hyper-multiplets in the $\CN=4$ theory take the same
representation of the gauge group. In this section, we will show
that this is also sufficient for enhancement to $\CN=5$. In other
words, for {\em any} (extended) $\CN=4$ Gaiotto-Witten theory, if
the two types of hyper-multiplets are in the same representation
of the gauge group so that $t^m_{AB}=\tilde{t}^m_{AB}$, the
supersymmetry is {\em automatically} enhanced to $\CN=5$.

The lift from $\CN=4$ to $\CN=5$ is an exercise of embedding the
$R$-symmetry group $SO(4)=SU(2)_L \times SU(2)_R = Sp(2)\times
Sp(2)$ into $Sp(4)=SO(5)$ in the standard way. We combine the
$\CN=4$ hyper and twisted hyper-multiplets to form $\CN=5$
multiplets,
\be
\Phi_\a^A =
\begin{pmatrix}
q_\a^A \\ \tilde{q}_\da^A
\end{pmatrix}\, ,
\;\;\;\;\;
\Psi_\a^A =
\begin{pmatrix}
\tilde{\psi}_\a^A \\ \psi_\da^A
\end{pmatrix} .
\ee
The reality conditions can be rewritten in the $\CN=5$ covariant way as
\be
 \bar \Phi_A^\a=(\Phi^A_\a)^\dagger=C^{\a\b}\omega_{AB}\Phi^B_\b\,,~~~~~~
 \bar \Psi_\a^A=(\Psi^A_\a)^\dagger=C^{\a\b}\omega_{AB}\Psi^B_\b\,,
\label{real5} \ee
where the invariant tensor of $Sp(4)$,
\be
C^{\a\b} =
\begin{pmatrix}
\e^{\a\b} & 0 \\ 0 & \e^{\da\db}
\end{pmatrix}\, ,
\ee
can be understood as the charge conjugation matrix for
the $SO(5)$ Clifford algebra in a suitably chosen basis.
The ``moment map'' and the ``current'' operators also take the
$\CN=5$ form,
 \be \CM^m_{\a\b} \equiv t^m_{AB} \Phi^A_\a \Phi^B_\b
\,, \;\;\;\;\; \CJ^m_{\a\b} \equiv t^m_{AB} \Phi^A_\a \Psi^B_\b
\,. \ee
After a slightly lengthy algebra (see appendix \ref{alg}), we can
uplift  the $\CN=4$ Lagrangian (\ref{Lful4}) in the $\CN=5$
Lagrangian,
\begin{eqnarray}
 {\cal L} &=&
 \frac{\varepsilon^{\m\n\l}}{4\pi}
 \left(k_{mn} A^m_\m \partial_\n A^n_\l
      +\frac13 f_{mnp} A^m_\m A^n_\n A^p_\l \right)
\nn\\&&
 +\frac12\omega_{AB} C^{\a\b}
  \left(-D \Phi^A_\a D \Phi^B_\b
        +i\Psi^A_\a \Ds\Psi^B_\b \right)
 -i\pi k_{mn} C^{\a\b} C^{\g\d}\left( \CJ^m_{\a\g}\CJ^n_{\b\d}
  -2\CJ^m_{\a\g}\CJ^n_{\d\b} \right)
    \nn\\&&
  +\frac{2\pi^2}{15}f_{mnp}(\CM^m)^\a_{~\b}(\CM^n)^\b_{~\g}(\CM^p)^\g_{~\a}
  +\frac{3\pi^2}{5}(\CM^{mn})^\g_{~\g}(\CM_m)^\a_{~\b}(\CM_n)^\b_{~\a} \,,
\label{Lful5}
\end{eqnarray}
and the supersymmetry transformation law,
\begin{eqnarray}
 && \delta \Phi_\alpha^A
    = i\eta_\alpha^{~\;\beta}\Psi_{\beta}^A\,,
~~~
    \delta A^m_\mu = 2\pi i\eta^{\a\b}\gamma_\mu \CJ^m_{\a\b}\,,
 \nn\\
 && \delta\Psi_{\alpha}^A
    = \left[\Ds \Phi_\g^A
           +\frac{2\pi}3(t_m)^A_{~B}\Phi^B_\beta(\CM^m)^\beta_{~\g}\right]
            \eta^\g_{~\;\alpha}
           -\frac{4\pi}{3} (t_m)^A_{~B}\Phi^B_\beta(\CM^m)^{\g}_{~\alpha}
            \eta^\beta_{~\;\g} \,.
\label{Sful5}
\end{eqnarray}
The parameter $\eta_{\a\b}$ satisfies
\be
\eta_{\a\b} = - \eta_{\b\a}\,,
\;\;\;\;\;
(\eta^*)^{\a\b} = - C^{\a\g}C^{\b\d} \eta_{\g\d}\,,
\;\;\;\;\;
C^{\a\b}\eta_{\a\b} =0 \,.
\ee

\subsection{$OSp(N|2M)$ example}

\paragraph{Symplectic embedding}

Let us denote the generators of $O(N)$ and $Sp(2M)$ as
$M_{ab}$ and $M_{\dot a\dot b}$, respectively.
The invariant anti-symmetric tensor of $Sp(2M)$ is denoted by
$\omega_{\dot a\dot b}$.
We denote the bi-fundamental matter fields $\Phi_\alpha^A,\Psi_\alpha^A$
by
\begin{equation}
 \Phi_\alpha^{a\dot a}, ~~~
 \Psi_\alpha^{a\dot a}.
\end{equation}
We choose the symplectic invariant tensor $\omega_{AB}$
as $\omega_{a\dot a,b\dot b}=\delta_{ab}\cdot\omega_{\dot a\dot b}$.
The matter fields obey the reality condition of the form
\begin{equation}
 \bar\Phi^\alpha_{\dot aa} ~=~
 (\Phi_\alpha^{a\dot a})^\dagger ~=~
 \delta_{ab}\omega_{\dot a\dot b}C^{\alpha\beta}\Phi_\beta^{b\dot b},
 \label{real5x}
\end{equation}
and similarly for the fermions.
In the following the $O(N)$ vector indices are raised or lowered
sloppily while the $Sp(2M)$ vector indices are raised or lowered by
$\omega_{\dot a\dot b}$ and $\omega^{\dot a\dot b}$.
Later we find it convenient to regard the matter fields as
$N\times 2M$ matrices and omit the indices.

{}From the commutation relation of $OSp(N|2M)$ generators,
\begin{eqnarray}
 ~[M_{ab},M_{cd}] &=&
  \delta_{bc}M_{ad}-\delta_{bd}M_{ac}
 -\delta_{ac}M_{bd}+\delta_{ad}M_{bc}\,,\nn\\
 ~[M_{\dda\ddb},M_{\ddc\ddd}] &=&
  \omega_{\ddb\ddc}M_{\dda\ddd} +\omega_{\ddb\ddd}M_{\dda\ddc}
 +\omega_{\dda\ddc}M_{\ddb\ddd} +\omega_{\dda\ddd}M_{\ddb\ddc}\,,\nn\\
 ~[M_{ab},Q_{c\ddc}] &=& \delta_{bc}Q_{a\ddc}-\delta_{ac}Q_{b\ddc}\,,\nn\\
 ~[M_{\dda\ddb},Q_{c\ddc}] &=&
  \omega_{\dda\ddc}Q_{c\ddb}+\omega_{\ddb\ddc}Q_{c\dda}\,,\nn\\
 ~\{Q_{a\dda},Q_{b\ddb}\} &=& \frac{k}{2\pi}
    (\omega_{\dda\ddb}M_{ab}+\delta_{ab}M_{\dda\ddb})\,,
\end{eqnarray}
one can read off the representation matrices on matters,
\begin{equation}
 (t_{ab})_{c\dot c, d\dot d}
 = \omega_{\dot c\dot d}(\delta_{ac}\delta_{bd}-\delta_{ad}\delta_{bc})\,,~~~~~
 (t_{\dot a\dot b})_{c\dot c, d\dot d}
 = -\delta_{cd}(\omega_{\dot a\dot c}\omega_{\dot b\dot d}
               +\omega_{\dot a\dot d}\omega_{\dot b\dot c})\, ,
\end{equation}
and the quadratic invariant tensor (Chern-Simons coupling)
\begin{equation}
 k^{ab,cd}~=~\frac{k}{8\pi}
 (\delta^{ac}\delta^{bd}-\delta^{ad}\delta^{bc})\,,~~~~~
 k^{\dot a\dot b,\dot c\dot d}~=~ -\frac k{8\pi}
 (\omega^{\dot a\dot c}\omega^{\dot b\dot d}
 +\omega^{\dot a\dot d}\omega^{\dot b\dot c}) \,.
\end{equation}

\paragraph{Lagrangian}
The kinetic terms for matters are given by
\begin{eqnarray}
  {\cal L}_{\text{kin}}
 = \frac12 \text{tr}\left(
   - D_\mu\bar\Phi^\alpha D^\mu\Phi_\alpha
   +i\bar\Psi^\alpha \sla D\Psi_\alpha \right)\, .
\end{eqnarray}
We normalize the gauge fields for each gauge group $O(N)$ and $Sp(2M)$ as
\begin{eqnarray}
  A_{O(N)}= \frac12 t_{ab}A^{ab}\,,~~~~~
  A_{Sp(2M)}= \frac12 t_{\dot a\dot b}
  (\omega^{\dot a\dot c}\tilde A_{\dot c}^{~\dot b})\,.
\end{eqnarray}
Then the Chern-Simons term becomes
\begin{eqnarray}
  {\cal L}_{\text{CS}} =
 \frac{\e^{\mu\nu\rho}}{4k}\text{tr}
  \left( - A_\mu \partial_\nu A_\rho
         - \frac{2}{3} A_\mu A_\nu A_\rho
         + \tilde{A}_\mu \partial_\nu \tilde{A}_\rho
         + \frac{2}{3} \tilde{A}_\mu \tilde{A}_\nu \tilde{A}_\rho \right)\ .
\end{eqnarray}
The Yukawa and bosonic potential terms are computed by
substituting the following expressions into the currents
and moment maps,
\be
 ({\cal J}_{ab})_{\alpha\beta}=
 (\Phi_\alpha\bar\Psi_\beta+\Psi_\beta\bar\Phi_\alpha)_{ab}\,,&&
 ({\cal J}_{\dot a\dot b})_{\alpha\beta}=
 (\bar\Phi_\alpha\Psi_\beta\omega
 +\bar\Psi_\beta\Phi_\alpha\omega)_{\dot a\dot b}\,,
\label{OSp-J}
\\
 ({\cal M}_{ab})_{\alpha\beta}=
 (\Phi_\alpha\bar\Phi_\beta+\Phi_\beta\bar\Phi_\alpha)_{ab}\,,&&
 ({\cal M}_{\dot a\dot b})_{\alpha\beta}=
 (\bar\Phi_\alpha\Phi_\beta\omega
 +\bar\Phi_\beta\Phi_\alpha\omega)_{\dot a\dot b}\,,
\label{OSp-M}
\ee
and so on.

For the computation of the interaction terms, it is useful to write
the currents (\ref{OSp-J}) and the moment maps (\ref{OSp-M}) into
the trace form,
\be
 ({\cal J}_{ab})_{\alpha\beta}=
 -{\rm tr}\left[\bar\Psi_\beta\tau_{ab}\Phi_\alpha\right],&&
 ({\cal J}_{\dot a\dot b})_{\alpha\beta}=
  {\rm tr}\left[\bar\Psi_\beta\Phi_\alpha\tau_{\dot a\dot b}\right],
\\
 ({\cal M}_{ab})_{\alpha\beta}=
 -{\rm tr}\left[\bar\Phi_\beta\tau_{ab}\Phi_\alpha\right],&&
 ({\cal M}_{\dot a\dot b})_{\alpha\beta}=
  {\rm tr}\left[\bar\Phi_\beta\Phi_\alpha \tau_{\dot a\dot b}\right],
\ee
Similarly, we also have
\begin{eqnarray}
 ({\cal M}_{ab,cd})_{\alpha\beta} &=&
 +{\rm tr}\left[\bar\Phi_\beta\tau_{cd}\tau_{ab}\Phi_\alpha\right], \nn\\
 ({\cal M}_{ab,\dot a\dot b})_{\alpha\beta} &=&
 -{\rm tr}\left[\bar\Phi_\beta\tau_{ab}
                \Phi_\alpha\tau_{\dot a\dot b}\right], \nn\\
 ({\cal M}_{\dot a\dot b,\dot c\dot d})_{\alpha\beta} &=&
 +{\rm tr}\left[\bar\Phi_\beta\Phi_\alpha
                \tau_{\dot a\dot b}\tau_{\dot c\dot d}\right].
\end{eqnarray}
Here $\tau_{ab}$ and $\tau_{\dot a\dot b}$ are the matrices in
the defining representation,
\begin{equation}
 (\tau_{ab})_{cd}=\delta_{ac}\delta_{bd}-\delta_{ad}\delta_{bc},~~~~
 (\tau_{\dot a\dot b})_{\dot c}^{~\dot d}=
 -\omega_{\dot a\dot c}\delta_{\dot b}^{\dot d}
 -\omega_{\dot b\dot c}\delta_{\dot a}^{\dot d}.
\end{equation}
Using the completeness relations, we can rewrite the product of
traces into a single trace,
\begin{eqnarray}
 k^{ab,cd}
 {\rm tr}\left[X\tau_{ab}\right]
 {\rm tr}\left[Y\tau_{cd}\right]
 &=& -\frac k\pi{\rm tr}\left[X_-Y_-\right],\nn\\
 k^{\dot a\dot b,\dot c\dot d}
 {\rm tr}\left[X\tau_{\dot a\dot b}\right]
 {\rm tr}\left[Y\tau_{\dot c\dot d}\right]
 &=& +\frac k\pi{\rm tr}\left[X_+Y_+\right],
\end{eqnarray}
where $X_-$ and $X_+$ are the projections of $X$ satisfying
$X^T=-X$ and $(X\omega)^T=+X\omega $.

Some straightforward computations give the Yukawa Lagrangian,
\begin{eqnarray}
 {\cal L}_{\rm Yukawa}
   &=&
  +\frac{ik}2{\rm tr}\left[
  -\bar\Psi_\beta\Phi_\alpha\bar\Phi^\alpha\Psi^\beta
  +\Psi_\beta\bar\Phi_\alpha\Phi^\alpha\bar\Psi^\beta
  +2\bar\Psi_\alpha\Phi_\beta\bar\Phi^\alpha\Psi^\beta
  -2\Psi^\beta\bar\Phi^\alpha\Phi_\beta\bar\Psi_\alpha
 \right]
 \nn\\ &&
  -ik\epsilon^{\alpha\beta\gamma\delta}
  {\rm tr}\left[\Phi_\alpha\bar\Psi_\beta\Phi_\gamma\bar\Psi_\delta\right] \,,
\end{eqnarray}
and the bosonic potential,
\begin{eqnarray}
 -V &=&
 \frac{k^2}{6}
  {\rm tr} \lbrace
   \Phi_\alpha\bar\Phi^\beta
   \Phi_\beta \bar\Phi^\gamma
   \Phi_\gamma\bar\Phi^\alpha
 + \Phi_\alpha\bar\Phi^\alpha
   \Phi_\beta \bar\Phi^\beta
   \Phi_\gamma\bar\Phi^\gamma
 \nn\\&& ~~~~~~~~
+ 4\Phi_\beta \bar\Phi^\alpha
   \Phi_\gamma\bar\Phi^\beta
   \Phi_\alpha\bar\Phi^\gamma
- 6\Phi_\gamma\bar\Phi^\gamma\Phi_\beta
   \bar\Phi^\alpha\Phi_\alpha\bar\Phi^\beta
 \rbrace ~.
\end{eqnarray}
In the computation of these we used the $Sp(4)$ identities
\be
\e^{\a\b\g\d} &=& C^{\a\b}C^{\g\d} + C^{\a\g} C^{\d\b}+ C^{\a\d} C^{\b\g} \,,
\label{id56}
\\
 \epsilon^{\alpha\beta\gamma\delta}\epsilon_{\alpha\rho\sigma\tau}
 &=& 6\delta^{[\beta}_\rho\delta^{\gamma}_\sigma\delta^{\delta]}_\tau
 \;=\; -3\left(
    \delta^\beta _\rho C^{\gamma\delta}C_{\sigma\tau}
 +  \delta^\gamma_\rho C^{\delta\beta }C_{\sigma\tau}
 +  \delta^\delta_\rho C^{\beta \gamma}C_{\sigma\tau}
       \right) \,,
\label{id56-2}
\ee
and an equality which follows directly from (\ref{id56-2}),
\begin{eqnarray}
 0 &=&
  {\rm tr} \lbrace
 -\Phi_\alpha\bar\Phi^\beta
  \Phi_\beta \bar\Phi^\gamma
  \Phi_\gamma\bar\Phi^\alpha
 -\Phi_\alpha\bar\Phi^\alpha
  \Phi_\beta \bar\Phi^\beta
  \Phi_\gamma\bar\Phi^\gamma
 -\Phi_\beta \bar\Phi^\alpha
  \Phi_\gamma\bar\Phi^\beta
  \Phi_\alpha\bar\Phi^\gamma
+3\Phi_\gamma\bar\Phi^\gamma\Phi_\beta
  \bar\Phi^\alpha\Phi_\alpha\bar\Phi^\beta
 \nn\\&&
+3\Phi_\alpha\bar\Phi^\alpha\Phi_\beta \bar\Phi_\gamma
  \Phi^\beta\bar\Phi^\gamma
-3\bar\Phi_\alpha \Phi^\alpha\bar\Phi_\gamma\Phi_\beta
  \bar\Phi^\gamma\Phi^\beta
+3\Phi^\alpha\bar\Phi^\beta\Phi^\gamma\bar\Phi_\alpha
  \Phi_\gamma\bar\Phi_\beta
  \rbrace~.
\end{eqnarray}
%
In summary, the full Lagrangian for the $O(N)\times Sp(2M)$ theory is
\begin{eqnarray}
 {\cal L} &=&
 \frac{\e^{\mu\nu\rho}}{4k}\text{tr}
  \left( - A_\mu \partial_\nu A_\rho
         - \frac{2}{3} A_\mu A_\nu A_\rho
         + \tilde{A}_\mu \partial_\nu \tilde{A}_\rho
         + \frac{2}{3} \tilde{A}_\mu \tilde{A}_\nu \tilde{A}_\rho \right)
 \nn\\&&
  + \frac12 \text{tr}\left(
   - D_\mu\bar\Phi^\alpha D^\mu\Phi_\alpha
   +i\bar\Psi^\alpha \sla D\Psi_\alpha \right)
  -ik\epsilon^{\alpha\beta\gamma\delta}
  {\rm Tr}\left(
     \Phi_\alpha\bar\Psi_\beta\Phi_\gamma\bar\Psi_\delta
  \right)
 \nn\\&&
  +\frac{ik}2{\rm tr}\left(
  -\bar\Psi_\beta\Phi_\alpha\bar\Phi^\alpha\Psi^\beta
  +\Psi_\beta\bar\Phi_\alpha\Phi^\alpha\bar\Psi^\beta
  +2\bar\Psi_\alpha\Phi_\beta\bar\Phi^\alpha\Psi^\beta
  -2\Psi^\beta\bar\Phi^\alpha\Phi_\beta\bar\Psi_\alpha
  \right)
 \nn\\ &&
 +\frac{k^2}{6}
  {\rm tr} \left(
   \Phi_\alpha\bar\Phi^\beta
   \Phi_\beta \bar\Phi^\gamma
   \Phi_\gamma\bar\Phi^\alpha
 + \Phi_\alpha\bar\Phi^\alpha
   \Phi_\beta \bar\Phi^\beta
   \Phi_\gamma\bar\Phi^\gamma
 \right.
 \nn\\&& ~~~~~~~~\left.
+ 4\Phi_\beta \bar\Phi^\alpha
   \Phi_\gamma\bar\Phi^\beta
   \Phi_\alpha\bar\Phi^\gamma
- 6\Phi_\gamma\bar\Phi^\gamma\Phi_\beta
   \bar\Phi^\alpha\Phi_\alpha\bar\Phi^\beta
   \right) \,.
   \label{Lful-OSp}
\end{eqnarray}

\paragraph{Supersymmetry transformation rules}
The $\CN=5$ supersymmetry transformation rule for
$OSp(N|2M)$ model is given by
\begin{eqnarray}
&& \delta \Phi_\alpha
    = i\eta_\alpha^{~\;\beta}\Psi_{\beta}\,,
\nn\\ &&
 \delta A_\mu
 =
 \frac{ik}2\eta^{\a\b}\gamma_\mu
 (\Phi_\alpha\bar\Psi_\beta+\Psi_\beta\bar\Phi_\alpha)\,,
 ~~~
 \delta\tilde A_\mu
 =
 \frac{ik}2\eta^{\a\b}\gamma_\mu
 (\bar\Phi_\alpha\Psi_\beta+\bar\Psi_\beta\Phi_\alpha)\,,
 \nn\\
&&  \delta\Psi_\alpha
 =
 \sla D\Phi_\gamma\eta^\gamma_{~\alpha}
 +\frac{2k}3(
  \Phi_{[\gamma}\bar\Phi^\beta\Phi_{\beta]}
 +\Phi^\beta\bar\Phi_\gamma\Phi_\beta )\eta^\gamma_{~\alpha}
 -\frac{4k}3(\Phi_{[\alpha}\bar\Phi^\gamma\Phi_{\beta]}
            +\Phi^\gamma\bar\Phi_\alpha\Phi_\beta )\eta^\beta_{~\gamma} .
            \;\;\;\;
\end{eqnarray}

\section{$\CN=6$ Superconformal Theories}
\label{n6}

In general, the Gaiotto-Witten construction we reviewed in section
2 assumes that the matter fields form a pseudo-real representation
$(\CR)$ of the gauge group; see the reality conditions
(\ref{real4}). If $\CR$ can be decomposed into a complex
representation $(R)$ and its complex-conjugate representation
$(\bar{R})$, then the $\CN=5$ supersymmetry is further enhanced to
$\CN=6$.

\subsection{General construction}

The construction is an exercise of embedding the $R$-symmetry group
$SO(5)=Sp(4)$ into $SU(4)=SO(6)$. The $\CN=5$ fields are
decomposed into $\CN=6$ fields as \footnote{To avoid introducing
new set of indices in every page, we are recycling not only the
$\a,\b$ indices, but also the $A,B$ indices. They run from $1$ to
$2n$ in $\CN=5$ formulas, but $1$ to $n$ in $\CN=6$ formulas.
Hopefully, the context would make it clear which notation is being
used. } \be (\Phi^A_\a)_{\CN=5} =
\begin{pmatrix}
\Phi_\a^A \\ C_{\a\b}\bar{\Phi}^\b_A
\end{pmatrix} \;,
\;\;\;\;\; (\Psi^A_\a)_{\CN=5} =
\begin{pmatrix}
C_{\a\b} \Psi^{\b A} \\ -\bar{\Psi}_{\a A}
\end{pmatrix} \;.
\ee With the symplectic invariant tensor, \be (\w_{AB})_{\CN=5} =
\begin{pmatrix}
0 & \d_A{}^B \\ -\d^A{}_B & 0
\end{pmatrix} \;,
\ee the reality conditions reduce to \be (\Phi_\a^A)^\dagger =
\bar{\Phi}^\a_A , \;\;\;\;\; (\Psi^{\a A})^\dagger =
\bar{\Psi}_{\a A} , \ee which is consistent with the following
assignments we need for the lift to $\CN=6$. \be
\begin{array}{l|cccc}
  & \; \Phi_\a^A \;  & \;\bar{\Phi}^\a_A \; & \;\Psi^{\a A}\; & \;\bar{\Psi}_{\a A} \; \\ \hline
{\rm Gauge} & R & \bar{R} & R & \bar{R} \\
SO(6)_R  & \mathbf{4} & \mathbf{\bar{4}} & \mathbf{\bar{4}} &
\mathbf{4}
\end{array}
\ee
The gauge generators are written in a block diagonal form as
\be (t^A{}_B)_{\CN=5} =
\begin{pmatrix}
t^A{}_B & 0 \\ 0 & -t^B{}_A
\end{pmatrix} \,,
\;\;\; (t_{AB})_{\CN=5} = - \begin{pmatrix} 0 & t^B{}_A \\ t^A{}_B
& 0
\end{pmatrix} \, ,
\ee
and the fundamental identity in the $\CN=6$ notation reads \be
(t^m)^{A}_{~B} (t_m)^C_{~D} + (t^m)^{A}_{~D} (t_m)^C_{~B} = 0 \,.
\label{fund6} \ee The ``moment map'' and ``current'' operators
have the decomposition, \be (\CM^m)^\a_{~\b} &=& - (M^m)^\a_{~\b}
- C^{\a\d}C_{\b\g} (M^m){}^\g_{~\d} \,, \label{MmN=6}
\\
(\CM^{mn})^\a_{~\b} &=& - (M^{mn})^\a_{~\b} + C^{\a\d}C_{\b\g}
(M^{nm}){}^\g_{~\d} \,, \label{MmnN=6}
\\
(\CJ^m)_{\a\b} &=& (J^m)_{\a\b} - C_{\a\g}C_{\b\d}
(\bar{J}^m)^{\g\d} \,, \label{JN=6} \ee where we introduced the
$\CN=6$ covariant quantities \be &&(M^m)^\a_{~\b} \equiv
\bar{\Phi}^\a_A (t^m)^A_{~B} \Phi^B_\b , \;\;\; (M^{mn})^\a_{~\b}
\equiv \bar{\Phi}^\a_A (t^m t^n)^A_{~B} \Phi^B_\b\,,
\\
&&(J^m)_{\a\b} \equiv \Phi_\a^B (t^m)^A_{~B}\bar{\Psi}_{\b A} ,
\;\;\; (\bar{J}^m)^{\a\b} \equiv \bar{\Phi}^\a_A (t^m)^A_{~B}
\Psi^{\b B} . \ee

To rewrite the $\CN=5$ Lagrangian of the previous section in an
$\CN=6$ covariant form, we have to make sure that all references
to $C_{\a\b}$ disappear. Using the $Sp(4)$ identity (\ref{id56})
\[
C^{\a\b}
C^{\g\d} + C^{\a\g} C^{\d\b}+ C^{\a\d} C^{\b\g} = \e^{\a\b\g\d},
\]
we can remove all $C_{\a\b}$ at the expense of
introducing $\e_{\a\b\g\d}$ which survives the lift.
After a slightly lengthy algebra (see appendix \ref{alg}), we
obtain the $\CN=6$ lift of the $\CN=5$ Lagrangian,
\begin{eqnarray}
 {\cal L} &=&
 \frac{\varepsilon^{\m\n\l}}{4\pi}
 \left(k_{mn} A^m_\m \partial_\n A^n_\l
      +\frac13 f_{mnp} A^m_\m A^n_\n A^p_\l \right)
 - D \bar{\Phi}_A^\a  D \Phi^A_\a
        +i\bar{\Psi}_{\a A} \Ds \Psi^{\a A}
 \nn\\&&
 +i\pi \left[ 2(\bar{J}_m)^{\a\b} (J^m)_{\a\b}
   -4(\bar{J}_m)^{\a\b} (J^m)_{\b\a}
 + \e^{\a\b\g\d} (J_m)_{\a\b} (J^m)_{\g\d}
  + \e_{\a\b\g\d} (\bar{J}_m)^{\a\b} (\bar{J}^m)^{\g\d} \right]
    \nn\\&&
  - \frac{4\pi^2}{3} f_{mnp}(M^m)^\a_{~\b}(M^n)^\b_{~\g}(M^p)^\g_{~\a}
  +4\pi^2 (M^{mn})^\a_{~\b}(M_m)^\b_{~\g}(M_n)^\g_{~\a} .
  \label{Lful6}
\end{eqnarray}
and the supersymmetry transformation law,
\begin{eqnarray}
 \delta \Phi_\alpha^A
    &=& -i\eta_{\alpha\beta}\Psi^{A\beta},
~~~
    \delta A^m_\mu = 2\pi i \left( \eta^{\a\b}\gamma_\mu (J^m)_{\a\b}
    + \eta_{\a\b}\gamma_\mu (\bar{J}^m)^{\a\b} \right) ,
 \nn\\
 \delta\Psi^{A\a}
    &=& \left[\Ds \Phi_\g^A
           -\frac{2\pi}3(t_m)^A_{~B}\Phi^B_\beta(M^m)^\beta_{~\g}\right]
            \eta^{\g\a}
           +\frac{4\pi}{3} (t_m)^A_{~B}\Phi^B_\beta(M^m)^{\a}_{~\g}
            \eta^{\g\b}
            \nn \\
&& - \frac{2\pi}{3} \e^{\a\b\g\d} (t_m)^A{}_B \Phi^B_\b
(M^m)^\rho{}_\g  \eta_{\d\rho} \,. \label{Sful6}
\end{eqnarray}
The parameter $\eta_{\a\b}$ satisfies \be \eta_{\a\b} = -
\eta_{\b\a}\,, \;\;\;\;\; (\eta^*)^{\a\b} = \half \e^{\a\b\g\d}
\eta_{\g\d}\,. \ee

\subsection{$U(M|N)$ example}

\paragraph{Symplectic embedding}

Let us denote the $U(M)$ and $U(N)$ generators as
$M_{a\Ua}$ and $M_{\dda\Uda}$, respectively.
Here the indices without underlines indicate fundamental
representation while those with underlines indicate
anti-fundamental representation. For the present model, the
complex matter fields $\Phi^A_\a$ and $\Psi^{A\a}$ are described
as
\begin{eqnarray}
  \Phi^A_\a = (\Phi_\a)^{a\Uda}\,, \hspace{0.5cm}
  \Psi^{A\a} = (\Psi^\a)^{a\Uda}\,,
\end{eqnarray}
and their complex conjugate fields as
\begin{eqnarray}
  \bar{\Phi}_A^\a = (\bar{\Phi}^\a)^{\dda \Ua}\,, \hspace{0.5cm}
  \bar{\Psi}_{A\a} = (\bar{\Psi}_\a)^{\dda \Ua}\,.
\end{eqnarray}
Hereafter we omit the indices $a,\dda,\Ua,\Uda$ and regard
$\Phi_\a, \Psi^\a$ as $M\times N$ matrices. We choose the
symplectic invariant tensor as
\begin{eqnarray}
  \omega_{a\Uda,\ddb\Ub} = - \omega_{\ddb\Ub,a\Uda}
 = \d_{a\Ub}\d_{\ddb\Uda}\ .
\end{eqnarray}
{}From the commutation relation of the Lie super-algebra $U(M|N)$
\begin{eqnarray}
\begin{array}{rcl}
~ [M_{a\Ub},Q_{c\Udc}]&=&
   +\delta_{c\Ub}Q_{a\Udc},
 \nn\\
~ [M_{a\Ub},\bar Q_{\ddc\Uc}]&=&
   -\delta_{a\Uc}\bar Q_{\ddc\Ub},
\end{array}
~~
\begin{array}{rcl}
~ [M_{\dda\Udb},Q_{c\Udc}]&=&
   -\delta_{\dda\Udc}Q_{c\Udb},
 \nn\\
~ [M_{\dda\Udb},\bar Q_{\ddc\Uc}]&=&
   +\delta_{\ddc\Udb}\bar Q_{\dda\Uc},
\end{array}
~~
\begin{array}{rcl}
~ [M_{a\Ub},M_{c\Ud}]&=&
 \delta_{c\Ub}M_{a\Ud}-\delta_{a\Ud}M_{c\Ub},
 \nn\\
~ [M_{\dda\Udb},M_{\ddc\Udd}]&=&
  \delta_{\ddc\Udb}M_{\dda\Udd}-\delta_{\dda\Udd}M_{\ddc\Udb},
\end{array}
\end{eqnarray}
\begin{equation}
 \{Q_{a\Uda},\bar Q_{\ddb\Ub}\}= \frac{k}{2\pi}
 \left(
  \delta_{\ddb\Uda}M_{a\Ub}+\delta_{a\Ub}M_{\ddb\Uda}
\right),
\end{equation}
one reads off the representation matrix of gauge group on matters
\begin{eqnarray}
  (t_{a\Ub})_{c\Udc,\ddd\Ud}= - \d_{\ddd\Udc} \d_{c\Ub}\d_{a\Ud}\,, \hspace{0.5cm}
  (t_{\dda\Udb})_{c\Udc,\ddd\Ud}=  \d_{c\Ud} \d_{\dda\Udc}
  \d_{\ddd\Udb} \,,
\end{eqnarray}
and the quadratic invariant tensor
\begin{eqnarray}
  k^{a\Ub,c\Ud} = -\frac{k}{2\pi}\d^{a\Ud}\d^{c\Ub} \,, \hspace{0.5cm}
  k^{\dda\Udb, \ddc \Udd}
  = + \frac{k}{2\pi} \d^{\dda \Udd}\d^{\ddc \Udb}\, .
\end{eqnarray}

\paragraph{Lagrangian}
Once we normalize gauge fields for each gauge group $U(M)$ and $U(N)$ as
\begin{eqnarray}
  A_{U(M)}=t_{a\Ub} A^{a\Ub}, \hspace{0.5cm}
 \tilde{A}_{U(N)}= t_{\dda\Udb} \tilde A^{\dda\Udb},
\end{eqnarray}
it is easy to write down the Chern-Simons term and the
matter kinetic term in the matrix form.
To express the remaining interactions in terms of matrix fields
$\Phi$ and $\Psi$,
it is useful to write the currents and the moment maps in the
trace form,
\begin{eqnarray}
 (J_{a\Ub})_{\alpha\beta}
 ={\rm tr}[\Phi_\alpha\bar\Psi_\beta\tau_{a\Ub}],
&&
 (J_{\dda\Udb})_{\alpha\beta}
 =-{\rm tr}[\Phi_\alpha\tau_{\dda\Udb}\bar\Psi_\beta],
\nn\\
 (\bar J_{a\Ub})^{\alpha\beta}
 ={\rm tr}[\Psi^\beta\bar\Phi^\alpha\tau_{a\Ub}],
&&
 (\bar J_{\dda\Udb})^{\alpha\beta}
 =-{\rm tr}[\Psi^\beta\tau_{\dda\Udb}\bar\Phi^\alpha],
\nn\\
 (M_{a\Ub})^\alpha_{~\beta}
 ={\rm tr}[\Phi_\beta\bar\Phi^\alpha\tau_{a\Ub}],
&&
 (M_{\dda\Udb})^\alpha_{~\beta}
 =-{\rm tr}[\Phi_\beta\tau_{\dda\Udb}\bar\Phi^\alpha],
\end{eqnarray}
\begin{eqnarray}
 (M_{a\Ub, c\Ud})^\alpha_{~\beta}
 &=& +{\rm tr}[\Phi_\beta\bar\Phi^\alpha\tau_{a\Ub}\tau_{c\Ud}],\nn\\
 (M_{a\Ub,\dda\Udb})^\alpha_{~\beta}
 &=& -{\rm tr}[\Phi_\beta\tau_{\dda\Udb}\bar\Phi^\alpha\tau_{a\Ub}],\nn\\
 (M_{\dda\Udb,\ddc\Udd})^\alpha_{~\beta}
 &=& +{\rm tr}[\Phi_\beta\tau_{\ddc\Udd}\tau_{\dda\Udb}\bar\Phi^\alpha],
\end{eqnarray}
The products of traces can be simplified using completeness relations
\begin{equation}
 k^{a\Ub, c\Ud}{\rm tr}[X\tau_{a\Ub}]{\rm tr}[Y\tau_{c\Ud}]=
 -\frac{k}{2\pi}{\rm tr}[XY],~~~~
 k^{\dda\Udb, \ddc\Udd}{\rm tr}[X\tau_{\dda\Udb}]{\rm tr}[Y\tau_{\ddc\Udd}]=
  \frac{k}{2\pi}{\rm tr}[XY].
\end{equation}
To summarize, the full $\CN=6$ supersymmetric Lagrangian for the $U(M)\times U(N)$
theory is
\begin{eqnarray}
  {\cal L} &&\hspace{-0.2cm}
 = - \frac{\e^{\mu\nu\rho}}{2k}
     \text{tr} \left( A_\mu \partial_\nu A_\rho
    + \frac{2}{3} A_\mu A_\nu A_\rho
    - \tilde{A}_\mu \partial_\nu \tilde{A}_\rho
    - \frac{2}{3} \tilde{A}_\mu \tilde{A}_\nu \tilde{A}_\rho \right)
  \nonumber \\ && \hspace{-0.4cm}
    - \text{tr} \left( D_\mu \bar{\Phi}^\a D^\mu \Phi_\a
    - i \bar{\Psi}_\a \g^\mu D_\mu \Psi^\a \right)
    -ik \epsilon^{\a\b\g\d}
    \text{tr}\left(\Phi_\a \bar{\Psi}_\b \Phi_\g \bar{\Psi}_\d \right)
    +ik \epsilon_{\a\b\g\d}
    \text{tr}\left(\bar{\Phi}^\a \Psi^\b \bar{\Phi}^\g \Psi^d \right)
  \nonumber \\ && \hspace{-0.5cm}
    -ik \text{tr}\left( \bar{\Phi}^\a \Phi_\a \bar{\Psi}_\b \Psi^\b
        - \Phi_\a \bar{\Phi}^\a \Psi^\b \bar{\Psi}_\b
        +2 \bar{\Phi}^\a \Psi^\b \bar{\Psi}_\a \Phi_\b
        -2 \Phi_\a \bar{\Psi}_\b \Psi^\a \bar{\Phi}^\b \right)
  \nonumber \\ && \hspace{-0.4cm}
  + \frac{1}{3}k^2 \text{tr} \left(
       \Phi_\a \bar{\Phi}^\a \Phi_\b \bar{\Phi}^\b \Phi_\g \bar{\Phi}^\g
    + \bar{\Phi}^\a \Phi_\a \bar{\Phi}^\b \Phi_\b \bar{\Phi}^\g \Phi_\g
         \right)
  + \frac{4}{3} k^2 \text{tr}\left(
    \Phi_\a \bar{\Phi}^\g \Phi_\b \bar{\Phi}^\a \Phi_\g \bar{\Phi}^\b \right)
  \nonumber \\ && \hspace{-0.4cm}
   - 2k^2 \text{tr}\left(
    \Phi_\a \bar{\Phi}^\a \Phi_\b \bar{\Phi}^\g \Phi_\g \bar{\Phi}^\b \right)\ .
\end{eqnarray}
The bosonic part is precisely that of the ABJM model \cite{ABJM}
and the Yukawa term agrees with that obtained in Ref.~\cite{a2}. 
Note that the $\CN=6$ Lagrangian above looks almost identical 
to the $\CN=5$ Lagragian (\ref{Lful-OSp}) of the $OSp(N|2M)$ model, 
except for the reality condition (\ref{real5x}) for the latter. 
It follows that the moduli space of vacua of the $OSp(N|2M)$ theory 
should be that of the $U(N|2M)$ theory modded out 
by the reality condition. We will come back to this point in section 4.

\paragraph{Supersymmetry transformation rules}
For scalar and gauge fields, one find
\begin{eqnarray}
  \d \Phi_\a &&\hspace{-0.2cm} = - i \eta_{\a \b} \Psi^\b, \hspace{0.3cm}
 \d A_\mu = ik ( \eta^{\a\b} \g_\mu \Phi_\a \bar{\Psi}_\b
               + \eta_{\a\b} \g_\mu \Psi^\b \bar{\Phi}^\a ), \nonumber \\
 \d \tilde A_\mu &&\hspace{-0.2cm} =
  ik (\eta^{\a\b} \g_\mu \bar{\Psi}_\b \Phi_\a
    + \eta_{\a\b} \g_\mu \bar{\Phi}^\a \Psi^\b ) \ .
\end{eqnarray}
The supersymmetry transformation rule for fermions now becomes
\begin{eqnarray}
  \hspace{-1cm}
  \d \Psi^\a  = \left[ \g^\mu D_\mu \Phi_\g - \frac{2k}{3}(\Phi_{[\b} \bar{\Phi}^\b \Phi_{\g]} )  \right]\eta^{\g\a} + \frac{4k}{3} (\Phi_{\b} \bar{\Phi}^\a \Phi_{\g}) \eta^{\g\b} - \frac{2k}{3} \e^{\a\b\g\d} (\Phi_{\b}   \bar{\Phi}^\r \Phi_{\g}) \eta_{\d\r}\ ,
\end{eqnarray}
in agreement with a recent independent work \cite{ggy}.

\subsection{$OSp(2|2M)$ example}

We now describe new $\CN=6$ superconformal Chern-Simons
theories for the super-algebra $OSp(2|2M)$.
The commutation relations were already discussed in section 2.3.

\paragraph{Symplectic embedding}  $U(1)=SO(2)$
and $Sp(2M)$ generators are denoted by $M_{+-}$ and $M_{ab}$,
respectively. Matter fields $\Phi^A_\a$ and $\Psi^{A\a}$ are
denoted by
\begin{eqnarray}
  \Phi^A_\a = (\Phi_\a)^{+a}, \hspace{0.5cm}  \Psi^{A\a} = (\Psi^\a)^{+a},
\end{eqnarray}
and their complex conjugate fields by
\begin{eqnarray}
  \bar{\Phi}_A^\a = (\bar{\Phi}^\a)_{+a}, \hspace{0.5cm}  \bar{\Psi}_{A\a} = (\bar{\Psi}_\a)_{+a}\ .
\end{eqnarray}
For clarity, we hereafter suppress the $U(1)$ and symplectic
indices of the matter fields. We choose the symplectic invariant
tensor to be
\begin{eqnarray}
  \omega_{+ a, - b} = \omega_{-a.+b} = \omega_{ab}\ ,
\end{eqnarray}
and the representation of gauge group on matter fields to be
\begin{eqnarray}
  (t_{+-})_{-a,+b} && \hspace{-0.2cm}= - (t_{+-})_{+a,-b} = \omega_{ab}  \\
  (t_{ab})_{\pm c, \mp d} &&\hspace{-0.2cm} = - (\omega_{ac}\omega_{bd} + \omega_{ad} \omega_{bc})\ .
\end{eqnarray}
The canonical expression used in section 3.1 can be obtained by
\begin{eqnarray}
  (t_m)^A_{\ B}: \ (t_{+-})^{+a}_{\ +b} && \hspace{-0.2cm} =\omega^{+a,-c}(t_{+-})_{-c,+b}=\d^a_{\ b}, \nonumber \\ (t_{ab})^{+c}_{\ +d} && \hspace{-0.2cm}  =\omega^{+c,-f}(t_{ab})_{-f,+d}=\d^c_{\ a}\omega_{bd} + \d^c_{\ b} \omega_{ad} \ .
\end{eqnarray}
The quadratic invariant tensor for $OSp(2|M)$ reads
\begin{eqnarray}
  k^{+-,+-} = -\frac{k}{2\pi}, \hspace{0.5cm} k^{ab,cd}= -\frac{k}{8\pi} \left( \omega^{ac}\omega^{bd} + \omega^{ad} \omega^{bc} \right)\ .
\end{eqnarray}
Before closing this paragraph, let us present some useful
quantities to be used below,
\begin{eqnarray}
   && (t_{+-}t_{+-})_{\pm a, \mp b} = \omega_{ab}, \nonumber \\ && (t_{ab}t_{cd})_{\pm f, \mp g} = - \left( \omega_{af} \omega_{bc} \omega_{dg} + \omega_{bf}\omega_{ac}\omega_{dg}+\omega_{af}\omega_{bd}\omega_{cg} +\omega_{bf}\omega_{ad}\omega_{cg}\right) \nonumber \\
   && (t_{+-}t_{ab})_{\pm c, \mp d}  = (t_{ab}t_{+-})_{\pm c, \mp d} =\pm \left(\omega_{ac}\omega_{bd} + \omega_{ad}\omega_{bc}\right)\ .
\end{eqnarray}

\paragraph{Lagrangian} In our convention, we obtain the kinetic terms for matter fields
\begin{eqnarray}
  {\cal L}_{\text{kin}} = -   D^\mu \Phi_\a D_\mu \bar{\Phi}^\a + i \Psi^\a \g^\mu D_\mu \bar{\Psi}_\a \ ,
\end{eqnarray}
together with the Chern-Simons term
\begin{eqnarray}
  {\cal L}_{\text{CS}} = - \frac{\e^{\mu\nu\rho}} {2k} A_\mu \partial_\nu A_\rho  + \frac{\e^{\mu\nu\rho}}{4k} \text{tr} \left(  \tilde{A}_\mu \partial_\nu \tilde{A}_\rho + \frac{2}{3} \tilde{A}_\mu \tilde{A}_\nu \tilde{A}_\rho \right)\ .
\end{eqnarray}
Here we normalized the gauge fields as
\begin{eqnarray}
  A_{U(1)}=t_{+-} A, \hspace{0.5cm} \tilde{A}_{Sp(2M)}= \frac12 t_{ab} \tilde{A}^{ab},
\end{eqnarray}
As usual, explicit expressions of the Yukawa interactions and
scalar potentials can be easily computed once we substitute the
moment map and current operators for the present model,
\begin{eqnarray}
  (M_{+-})^\da_{\ \db}= + (\Phi_\b \bar{\Phi}^\a), \hspace{0.5cm} (M_{ab})^\a_{\ \b} = + (\bar{\Phi}^\a \Phi_\b)_{ab} + (\bar{\Phi}^\a \Phi_\b )_{ba}\ ,
\end{eqnarray}
and
\begin{eqnarray}
  (J_{+-})_{\a\b} && \hspace{-0.2cm} = + (\Phi_\a \bar{\Psi}_\b), \hspace{0.3cm} (J_{ab})_{\a\b}  = + (\bar{\Psi}_\b \Phi_\a)_{ab} + (\bar{\Psi}_\b \Phi_\a)_{ba} , \nonumber \\
  (\bar{J}_{+-})^{\a\b}&& \hspace{-0.2cm} = + (\Psi^\b \bar{\Phi}^\a)\
  ,\hspace{0.15cm} (\bar{J}_{ab})^{\a\b} = + (\bar{\Phi}^\a \Psi^\b)_{ab} + (\bar{\Phi}^\a \Psi^\b)_{ba} \ .
\end{eqnarray}
As for the symplectic summation convention, we take $\Phi_1 \bar{\Phi}_2 \equiv \Phi_1^a \bar{\Phi}_{2a}$. 

In summary, the full Lagrangian is ${\cal L}={\cal L}_{\text{kin}}+{\cal L}_{\text{CS}}+{\cal
L}_{\text{Yukawa}}+{\cal L}_{\text{potential}}$, where
\begin{eqnarray}
  {\cal L}_{\text{kin}} && \hspace{-0.2cm} = -   D^\mu \Phi_\a D_\mu \bar{\Phi}^\a + i \Psi^\a \g^\mu D_\mu \bar{\Psi}_\a , \nonumber \\
  {\cal L}_{\text{CS}} && \hspace{-0.2cm}= - \frac{\e^{\mu\nu\rho}} {2k} A_\mu \partial_\nu A_\rho  + \frac{\e^{\mu\nu\rho}}{4k} \text{tr} \left(  \tilde{A}_\mu \partial_\nu \tilde{A}_\rho + \frac{2}{3} \tilde{A}_\mu \tilde{A}_\nu \tilde{A}_\rho \right),
  \nonumber \\
  {\cal L}_{\text{Yukawa}} && \hspace{-0.2cm} = -ik \left(\Phi_\a \bar{\Psi}_\b \cdot \Psi^\b \bar{\Phi}^\a - \Phi_\a \bar{\Phi}^\a \cdot \Psi^\b \bar{\Psi}_\b - \Phi_\a \omega \Psi^\b \cdot \bar{\Psi}_\b \omega \bar{\Phi}^\a\right)  \nonumber \\
  && \hspace{0.2cm} + 2ik \left(\Phi_\a \bar{\Psi}_\b \cdot \Psi^\a \bar{\Phi}^\b - \Phi_\a\bar{\Phi}^\b \cdot \Psi^\a \bar{\Psi}_\b - \Phi_\a \omega \Psi^\a \cdot \bar{\Psi}_\b \omega \bar{\Phi}^\b\right)
  \nonumber \\
  && \hspace{0.2cm} -i k \epsilon^{\a\b\g\d}
  \left(\Phi_\a \bar{\Psi}_\b \cdot \Phi_\g \bar{\Psi}_\d - \frac12 \Phi_\a \omega \Phi_\g \cdot \bar{\Psi}_\b \omega \bar{\Psi}_\d \right) \nonumber \\ && \hspace{0.2cm} - ik \epsilon_{\a\b\g\d}
  \left(\Psi^\b \bar{\Phi}^\a \cdot \Psi^\d \bar{\Phi}^\g - \frac12 \bar{\Phi}^\a \omega \bar{\Phi}^\g \cdot \Psi^\b \omega \Psi^\d  \right),
\nonumber \\
  {\cal L}_{\text{potential}} &&\hspace{-0.2cm} = - 3k^2 \left( \bar{\Phi}^\a \omega \bar{\Phi}^\b \cdot \Phi_\b \bar{\Phi}^\g \cdot \Phi_\g \omega \Phi_\a\right) + \frac{5k^2}{3} \left( \Phi_\a \bar{\Phi}^\b \cdot \Phi_\b \bar{\Phi}^\g \cdot \Phi_\g \bar{\Phi}^\a \right) \nonumber \\ && \hspace{0.2cm} - 2k^2 \left( \Phi_\a \bar{\Phi}^{\g} \cdot \Phi_\b \bar{\Phi}^\b \cdot \Phi_\g \bar{\Phi}^\a \right) + \frac{k^2}{3} \left( \Phi_\a \bar{\Phi}^\a \cdot \Phi_\b \bar{\Phi}^\b \cdot \Phi_\g \bar{\Phi}^\g \right)\ .
\end{eqnarray}
Here we used the notations
\begin{eqnarray}
  \Phi_1 \omega \Phi_2 = \Phi_1^a \omega_{ab} \Phi_2^b, \hspace{0.5cm} \bar{\Phi}_1 \omega \bar{\Phi}_2 = \bar{\Phi}_{1a} \omega^{ab} \bar{\Phi}_{2b}\ .
\end{eqnarray}

\paragraph{Supersymmetry transformation rules} The $\CN=6$ supersymmetry transformation rules for $OSp(2|M)$ model are given by
\begin{eqnarray}
  \d \Phi_\a &&\hspace{-0.2cm} = - i \eta_{\a \b} \Psi^\b, \hspace{0.3cm} \d A_\mu = -ik ( \eta^{\a\b} \g_\mu \Phi_\a \bar{\Psi}_\b + \eta_{\a\b} \g_\mu \Psi^\b \bar{\Phi}^\a) , \nonumber \\ \d A^{ab}_\mu &&\hspace{-0.2cm} = - ik (\eta^{\a\b} \g_\mu \bar{\Psi}_\b \Phi_\a + \eta_{\a\b} \g_\mu \bar{\Phi}^\a \Psi^\b)^{(ab)} \ .
\end{eqnarray}
and
\begin{eqnarray}
  \d \Psi^\a && \hspace{-0.2cm} = \left( \g^\mu D_\mu \Phi_\g - \frac{2k}{3} \Phi_{[\b} \bar{\Phi}^\b \cdot \Phi_{\g]} + \frac{k}{3} \omega \bar{\Phi}^\b \cdot \Phi_\b \omega \Phi_\g \right) \eta^{\g\a} \nonumber \\  &&
 \hspace{0.2cm} + \left( \frac{4k}{3} \Phi_{\b} \bar{\Phi}^\a \cdot \Phi_{\g} - \frac{2k}{3} \omega \bar{\Phi}^\a \cdot \Phi_\b \omega \Phi_\g \right) \eta^{\g\b}\nonumber \\&& \hspace{0.2cm} +  \e^{\a\b\g\d } \left( \frac{2k}{3}\Phi_{\b} \bar{\Phi}^\r \cdot \Phi_{\g} - \frac{k}{3} \omega \bar{\Phi}^\r \cdot \Phi_\b \omega \Phi_\g \right) \eta_{\d\r} \ .
\end{eqnarray}

\section{IIB Orientifold and M2-branes on Orbifold}
\label{ori}

In \cite{ABJM} it was argued that the ${\cal N}=6$ $U(N)\times
U(N)$ ABJM model with CS coupling $k$ is the world-volume theory of
$N$ M2-branes in orbifold $\CC^4/\ZZ_k$.
\footnote{To avoid confusion, 
note that the letter ``$k$'' used in earlier sections (2.3, 3.2, 3.3) 
of this paper is inversely related to the integer-quantized Chern-Simons level $k$ of this section.} 
 Here we argue that our
${\cal N}=5$ theory with the gauge group $SO(2N)\times Sp(2N)$ and
the CS coupling $2k$ is the world-volume theory of $N$ M2-branes
in orbifold $\CC^4/\hat D_{k+2}$, where $\hat D_k$ is the binary
dihedral group.

Our arguments closely follow that of \cite{ABJM}. We take the
orientifold of a Type IIB brane configuration realizing the ABJM
model, and consider its M-theory dual. We show how the orientifold
breaks supersymmetry down to  ${\cal N}=5$ from the viewpoint of
M-theory geometry as well as the world-volume field theory.

\paragraph{Brane construction of ABJM theory}

The ABJM model with gauge group $U(N)\times U(N)$ can be embedded
in IIB superstring theory in flat spacetime with compact $x_6$
direction. Consider $N$ D3-branes(0126) intersecting with an
NS5-brane(012345) and an $(1,k)$ 5-brane(0123'4'5') at different
points on the $S^1(x_6)$. The directions $3',4',5'$ are given by
rotating $3,4,5$ by the same angle $\theta$ in the planes $37, 48$
and $59$ respectively. The D3-brane world-volume theory is a ${\cal
N}=3$ $U(N)\times U(N)$ Yang-Mills Chern-Simons theory which flows
to the ABJM model in the IR limit.

T-duality along the $x_6$ direction followed by an M-theory lift
gives a theory of $N$ M2-branes. The transverse space is a
fibration of $T^2(\tilde{x}_6,x_{10})$ over
$\RR^6(x_3,x_4,x_5,x_{3'},x_{4'},x_{5'})$. The 5-branes turn into
Taub-NUT type geometry after the duality chain. The M-theory
geometry is given by a $\ZZ_k$ orbifold of the product of two
Taub-NUTs, where the $\ZZ_k$ is the simultaneous translation along
the two $S^1$ fibers by $1/k$-period. Thus the ABJM model
describes $N$ M2-branes at the orbifold $\CC^4/\ZZ_k$.

The four bi-fundamental scalars in ABJM model, which we denote by
$(\phi_1,\phi_2,\phi_3,\phi_4)$ here, are identified with complex
coordinates of the orbifold. As an example, for $N=1$ the scalars
$(\phi_i)$ are complex numbers. The $U(1)\times U(1)$ gauge fields
removes one dimension of the moduli space through gauge
equivalence and adds one back through the dual photon. The net
effect is the $\ZZ_k$ orbifolding. \be
 \alpha~:~\phi_i ~\longrightarrow~ e^{2\pi i/k}\phi_i.
\ee The $SU(4)$ $R$-symmetry of ABJM model has a geometric
interpretation as the subgroup of transverse $SO(8)$ rotations
which commutes with the $\ZZ_k$ orbifolding.

\paragraph{Introduction of orientifold}

Back in Type IIB setup, introducing the O3-plane on top of $2N$
D3-branes gives an $SO(2N)\times Sp(2N)$ gauge theory
\cite{Elitzur:1998ju}. Following the chain of duality to M-theory,
one finds that the orientifold turns into an orbifold $(\beta)$
that flips all the coordinates ($3,4,5;3',4',5';\tilde{6},10$).
Since the directions $\tilde{6}$ and 10 make the phase directions
of the fields $\phi_i$ and $\b$ reverses them, $\beta$ should act
anti-holomorphically on these fields. Also, if one requires that
the origin is the only fixed point under $\beta$, the action
cannot be involutive.

Let us recall the simpler system of O6$^-$-plane and $k$ D6-branes
that uplifts to the M-theory on $\hat D_{k+2}$ orbifold. The
generators $\alpha,\beta$ of the orbifold group $\hat D_{k+2}$
correspond to the $1/2k$-period shift along the M-theory circle
and the orientifold, respectively. They satisfy \be
 \alpha^{2k}=1,~~~~ \beta^2=\alpha^k,~~~~
 \beta\alpha\beta^{-1}=\alpha^{-1}.
\ee
So $\beta$ squares to the half-period shift along the M-theory
circle. If the same rule applies to our case, then $\beta^2$
should flip the sign of all the fields $\phi_i$. {}From
anti-holomorphicity and $\beta^2=-1$, the action of $\beta$ on
fields should be of this form
\be
 \beta~:~ (\phi_1,\phi_2,\phi_3,\phi_4)
 ~\longrightarrow~
 (\phi_2^*,-\phi_1^*,\phi_4^*,-\phi_3^*),
 \label{kazuo}
\ee up to a linear redefinition of fields. This is equivalent to
the reality condition of matter fields in ${\cal N}=5$
supersymmetric theory of section \ref{n5} involving the matrix
$C_{\alpha\beta}$. This new orbifold element breaks the transverse
rotation symmetry further to $Sp(4)\simeq SO(5)$, in consistency
with the supersymmetry of the world-volume theory.

\paragraph{Orientifolding the field theory}

As noted in Ref.~\cite{ABJM}, the ABJM model written in $d=3$,
$\CN=2$ super-field notation closely resembles the conifold theory
\cite{kw} in $d=4$, $\CN=1$ notation. The reason is that both
theories have dual descriptions in terms of D-branes winding
around a circle and intersecting with two 5-branes at different
points on the circle. Also, the two 5-branes are tilted relative
to each other in both theories, albeit in somewhat different ways.

Here we argue that the orientifold action (\ref{kazuo}) can be
obtained by the standard open string analysis. The bi-fundamental
matter fields of ABJM model arise from the open string connecting
a pair of D-branes separated by a 5-brane. The orientifold
projection on those fields should be independent of the relative
angle of the two 5-branes.

We start with the conifold theory described as $U(2N) \times
U(2N)$ gauge theory with two bi-fundamental fields $A_i \,\,\, (2N,
\overline{2N})$ and $B_i \,\,\, (\overline{2N},2N)$ and the
super-potential
\begin{equation}
W=\Tr(A_1B_2A_2B_1-A_1B_1A_2B_2). \label{supercon}
\end{equation}
One encounters the same super-potential when writing the ABJM model
in $d=3, {\cal N}=2$ chiral super-fields $A_i,B_i$. Their lowest
components are combined into an $SU(4)$ multiplet, \be
\Phi=\begin{pmatrix}
A_i \\
B_i^{\dagger}
\end{pmatrix}, \;\;\;\;\;
\bar{\Phi}=\begin{pmatrix}
(A^\dagger)^i \\
 B^i
\end{pmatrix} \,.
\label{jaemo} \ee The orientifold of the conifold model relevant
for our discussion was discussed in section 4.3.2 of
\cite{dimerori}. There it was shown that the orientifold acts on
the matter fields as the $\ZZ_2$ identification
\begin{equation}
A_1=  B_2^T J\equiv  A, \,\,\, A_2=- B_1^T J\equiv B.
\label{oriaction}
\end{equation}
where $J$ is a matrix form of the anti-symmetric invariant tensor
of $Sp(2N)$ satisfying $J^2=-1$. The reason why we should impose
the condition $J^2=-1$ instead of $J^2=1$ on the Chan-Paton
factors is explained at \cite{polchinskiz2}. The resulting theory
is an $SO(2N) \times Sp(2N)$ gauge theory with two bi-fundamental
fields $A,B$ and the super-potential
\begin{equation}
W=\Tr (AB^TBA^T-BB^TAA^T). \label{superori}
\end{equation}
One can see that (\ref{oriaction}) is nothing but the reality
condition on the field $\Phi$ of section \ref{n5} up to a trivial
change of basis.


\vspace{0.5cm}

\section*{Acknowledgments}
We thank Nakwoo Kim and  David Tong and for discussions.
K.M.L. and J.P. are  supported in part by the KOSEF SRC Program
through CQUeST at Sogang University. K.M.L. is supported in part
by KRF Grant No. KRF-2005-070-C00030, and the KRF National Scholar
program. Sm.L. is supported in part by the KOSEF Grant
R01-2006-000-10965-0 and the Korea Research Foundation Grant
KRF-2007-331-C00073.
K.H. acknowledges the hospitality of CEA Saclay where parts of
the work were carried out.
Sm.L. acknowledges the hospitality of Simons Workshop 2008 
where the paper was finalized. 
J.P. is supported in part  by the Stanford
Institute for Theoretical Physics. J.P. acknowledges the hospitality
of the Aspen Center for Physics
where parts of the project were done.

\newpage

\centerline{\large \bf Appendix}

\appendix

\section{Details of Computation}
\label{alg}

\subsection{The $\CN=5$ Case}

\paragraph{Yukawa term}

We start from the Yukawa terms of ${\cal N}=4$ Lagrangian
(\ref{Lful4}). By applying the fundamental identity for $t^m$ to
the last two terms we get,
\begin{eqnarray}
{\cal L}_{\rm Y} &=&
 i\pi\epsilon^{\alpha\beta}\epsilon^{\gamma\delta}
 \left\{
 -(q_\alpha t^m \psi_\gamma)(q_\beta t_m\psi_\delta)
 -(\tilde q_\alpha t^m \tilde\psi_\gamma)(\tilde q_\beta t_m\tilde\psi_\delta)
 +4(q_\alpha t^m \psi_\gamma)(\tilde q_\delta t_m\tilde\psi_\beta)
 \right.
 \nn\\&&~~~~~~~~~
 -(q_\alpha t^m\tilde\psi_\beta)(q_\gamma t_m\tilde\psi_\delta)
 -(\tilde q_\alpha t^m\psi_\beta)(\tilde q_\gamma t_m\psi_\delta)
 \nn\\&&
 \left.~~~~~~~~~
 -(q_\alpha t^m\tilde\psi_\delta)(q_\gamma t_m\tilde\psi_\beta)
 -(\tilde q_\alpha t^m\psi_\delta)(\tilde q_\gamma t_m\psi_\beta)
 \right\} \,. \nn
\end{eqnarray}
Here we dropped dots on indices as there is no confusion. Now we
replace the two terms in the second line of the RHS with similar
terms with $\beta,\gamma$ exchanged. This effect can be cancelled
by doubling the two terms in the third line once the
fundamental identities for $\epsilon^{\a\b}$ are used, and we get
the following
\begin{eqnarray}
{\cal L}_{\rm Y} &=&
 i\pi\epsilon^{\alpha\beta}\epsilon^{\gamma\delta}
 \left\{
 -(q_\alpha t^m \psi_\gamma)(q_\beta t_m\psi_\delta)
 -(\tilde q_\alpha t^m \tilde\psi_\gamma)(\tilde q_\beta t_m\tilde\psi_\delta)
 \right.
 \nn\\&&~~~~~~~~~
 -(q_\alpha t^m\tilde\psi_\gamma)(q_\beta t_m\tilde\psi_\delta)
 -(\tilde q_\alpha t^m\psi_\gamma)(\tilde q_\beta t_m\psi_\delta)
 \nn\\&&
 \left.~~~~~~~~~
 -2(q_\alpha t^m\tilde\psi_\delta)(q_\gamma t_m\tilde\psi_\beta)
 -2(\tilde q_\alpha t^m\psi_\delta)(\tilde q_\gamma t_m\psi_\beta)
 +4(q_\alpha t^m \psi_\gamma)(\tilde q_\delta t_m\tilde\psi_\beta)
 \right\} \nn \\ &=&
 -i\pi k_{mn}C^{\alpha\beta}C^{\gamma\delta}
  \left\{
    {\cal J}^m_{\alpha\gamma}{\cal J}^n_{\beta\delta}
  -2{\cal J}^m_{\alpha\gamma}{\cal J}^n_{\delta\beta}
  \right\}\,.
\end{eqnarray}

\paragraph{Potential}
We begin by collecting some useful formulae.
We start from
\begin{eqnarray}
 0 &=&
 (\mu^{mn})^\alpha_{~\beta}(\mu_m)^\beta_{~\gamma}(\mu_n)^\gamma_{~\alpha}
+(\mu^{mn})^\beta_{~\beta}(\mu_m)^\alpha_{~\gamma}(\mu_n)^\gamma_{~\alpha}
+(\mu^{mn})_{\gamma\beta}(\mu_m)^{\beta\alpha}(\mu_n)^\gamma_{~\alpha}
 \nn\\ &=&
2(\mu^{mn})^\alpha_{~\beta}(\mu_m)^\beta_{~\gamma}(\mu_n)^\gamma_{~\alpha}
+(\mu^{mn})^\beta_{~\beta}(\mu_m)^\alpha_{~\gamma}(\mu_n)^\gamma_{~\alpha}
 \nn\\ &=&
 f_{pmn}(\mu^p)^\alpha_{~\beta}(\mu^m)^\beta_{~\gamma}(\mu^n)^\gamma_{~\alpha}
+2(\mu^{mn})^\beta_{~\beta}(\mu_m)^\alpha_{~\gamma}(\mu_n)^\gamma_{~\alpha}\,.
\label{appN5-1}
\end{eqnarray}
Similar equalities hold if some $q$ are replaced by $\tilde q$,
which we express by putting dots to the indices.
Putting dots to $\beta$ we get
\begin{eqnarray}
 0 &=&
2(\mu^{mn})^\alpha_{~\dot\beta}
 (\mu_m)^{\dot\beta}_{~\gamma}
 (\mu_n)^\gamma_{~\alpha}
+(\mu^{mn})^{\dot\beta}_{~\dot\beta}
 (\mu_m)^\alpha_{~\gamma}
 (\mu_n)^\gamma_{~\alpha} \,.
\label{appN5-2}
\end{eqnarray}
Putting dots to $\alpha$ we get
\begin{eqnarray}
 0 &=&
 2(\mu^{mn})^{\dot\alpha}_{~\beta}
  (\mu_m)^\beta_{~\gamma}
  (\mu_n)^\gamma_{~\dot\alpha}
+2(\mu^{mn})^\beta_{~\beta}
  (\mu_m)^{\dot\alpha}_{~\gamma}
  (\mu_n)^\gamma_{~\dot\alpha}
+2(\mu^{mn})^\gamma_{~\beta}
  (\mu_m)^\beta_{~\dot\alpha}
  (\mu_n)^{\dot\alpha}_{~\gamma}
 \nn\\ &=&
-(\mu^{mn})^{\dot\beta}_{~\dot\beta}
 (\mu_m)^\alpha_{~\gamma}
 (\mu_n)^\gamma_{~\alpha}
+3(\mu^{mn})^\beta_{~\beta}
 (\mu_m)^{\dot\alpha}_{~\gamma}
 (\mu_n)^\gamma_{~\dot\alpha}
+f_{mnp}(\mu^m)^\gamma_{~\beta}
 (\mu^n)^\beta_{~\dot\alpha}
 (\mu^p)^{\dot\alpha}_{~\gamma} \,. \;\;
\label{appN5-3}
\end{eqnarray}
Here the first term was rewritten using
$(\mu^{mn})^\alpha_{~\dot\beta}=-(\mu^{nm})_{\dot\beta}^{~\alpha}$
and (\ref{appN5-2}), and the third term was decomposed into
symmetric and antisymmetric parts in $mn$.
Now consider
\begin{eqnarray}
 I &\equiv&
 2f_{mnp}
  ({\cal M}^m)^\alpha_{~\beta}
  ({\cal M}^n)^\beta_{~\gamma}
  ({\cal M}^p)^\gamma_{~\alpha}
 +9({\cal M}^{mn})^\alpha_{~\alpha}
  ({\cal M}_m)^\beta_{~\gamma}
  ({\cal M}_n)^\gamma_{~\beta} \,.
\end{eqnarray}
Expanding this into $\mu$'s and using (\ref{appN5-3}) and
(\ref{appN5-1}) we find
\begin{eqnarray}
 I &=&
 -\frac52f_{mnp}
  (\mu^m)^\alpha_{~\beta}
  (\mu^n)^\beta_{~\gamma}
  (\mu^p)^\gamma_{~\alpha}
 -\frac52f_{mnp}
  (\mu^m)^{\dot\alpha}_{~\dot\beta}
  (\mu^n)^{\dot\beta}_{~\dot\gamma}
  (\mu^p)^{\dot\gamma}_{~\dot\alpha}
 \nn\\&&
 +15(\mu^{mn})^{\dot\alpha}_{~\dot\alpha}
    (\mu_m)^{\beta}_{~\gamma}
    (\mu_n)^{\gamma}_{~\beta}
 +15(\mu^{mn})^{\alpha}_{~\alpha}
    (\mu_m)^{\dot\beta}_{~\dot\gamma}
    (\mu_n)^{\dot\gamma}_{~\dot\beta} \,.
\end{eqnarray}
Hence the potential term in (\ref{Lful4}) is
${\cal L}_{\rm pot}=-V=\pi^2 I/15$.

\paragraph{Supersymmetry transformation}

Let $\eta_{\alpha\dot\alpha}$ be the parameter of ${\cal N}=4$
supersymmetry.
We define $\eta_{\dot\alpha\alpha}=-\eta_{\alpha\dot\alpha}$, and
introduce the $4\times 4$ matrix valued spinor
\begin{equation}
 \hat\eta_\alpha^{~\beta}~=~\left(\begin{array}{cc}
   0 & \eta_\alpha^{~\dot\beta} \\  \eta_{\dot\alpha}^{~\beta} & 0
 \end{array}\right).
\end{equation}
Rewriting the ${\cal N}=4$ transformation law (\ref{Sful4}) in terms
of $Sp(4)$ multiplets $\Phi_\alpha^A,\Psi_\alpha^A$ and
$\hat\eta_\alpha^{~\beta}$, we easily obtain (\ref{Sful5}).

\paragraph{Sp(4) $R$-symmetry}
The $R$-symmetry acts on matter fields as
\begin{equation}
 {\Phi'}_\alpha^A~=~ U_\alpha^{~\beta}\Phi_\beta^A\,,~~~~~
 {\Psi'}_\alpha^A~=~ U_\alpha^{~\beta}\Psi_\beta^A\,.
\end{equation}
$U$ is unitary and satisfies $U^\ast=CUC^{-1}$, $U^T C U=C$.
They are equivalently $SO(5)$ spinors with charge conjugation matrix $C$.
The ${\cal N}=4$ supersymmetry parameter $\hat\eta$ satisfies,
\begin{equation}
 C\hat\eta C^{-1}=\hat\eta^T\,,~~~~
 (\hat\eta^\ast)=C\hat\eta C^{-1}\,,~~~~
 {\rm Tr}[\hat\eta]=0\,,~~~~
 \Gamma_5\hat\eta\Gamma_5=-\hat\eta\,,
\end{equation}
where $(\Gamma_5)_\alpha^{~\beta}={\rm diag}(+1,+1,-1,-1)$. The
$Sp(4)$ R-invariance removes the last condition and uplifts the
supersymmetry to ${\cal N}=5$.

\subsection{The $\CN=6$ Case}

\paragraph{Yukawa term}

Substituting (\ref{JN=6}) in the Yukawa terms of the ${\cal N}=5$
Lagrangian (\ref{Lful5}) and expanding, we get the ${\cal N}=6$ Yukawa
term ${\cal L}_{\rm Y}=i\pi I_{\rm Y}$.
\begin{eqnarray}
I_Y &=&
  2(J^n)_{\alpha\beta}(\bar J_n)^{\alpha\beta}
 -4(J^n)_{\alpha\beta}(\bar J_n)^{\beta\alpha}
 \nn\\&& ~ \hskip-6mm
 +(J^n)_{\alpha\beta}(J_n)_{\gamma\delta}(
  2C^{\alpha\delta}C^{\beta\gamma}-C^{\alpha\gamma}C^{\beta\delta})
 +(\bar J^n)^{\alpha\beta}(\bar J_n)^{\gamma\delta}(
  2C_{\alpha\delta}C_{\beta\gamma}-C_{\alpha\gamma}C_{\beta\delta}).\;\;\;
\end{eqnarray}
Using the fundamental identity (\ref{fund6}) and the $Sp(4)$
identity (\ref{id56}) we get the Yukawa terms in (\ref{Lful6}).

\paragraph{Potential}

The potential term is ${\cal L}_{\rm pot}=-V=\pi^2 I/15$, where
\be I = 2f_{mnp}(\CM^m)^\a_{~\b}(\CM^n)^\b_{~\g}(\CM^p)^\g_{~\a}
  + 9(\CM^{mn})^\g_{~\g}(\CM_m)^\a_{~\b}(\CM_n)^\b_{~\a} \equiv 2I_1 + 9I_2 \,.
\ee Substituting (\ref{MmN=6}) and (\ref{MmnN=6}) we obtain the
intermediate results, \be 2I_1 &=& -4 f_{mnp}
(M^m)^\a_{~\b}(M^n)^\b_{~\g}(M^p)^\g_{~\a} +12f_{mnp}
(M^m)^\a_{~\b}(M^n)^\b_{~\g} C^{\g\rho} (M^p)^{\s}_{~\rho}
C_{\s\a},
\nn \\
9I_2 &=& -36(M^{mn})^\g_{~\g} (M^m)^\a_{~\b}(M^n)^\b_{~\a}
  +36(M^{mn})^\g_{~\g} (M^m)^\a_{~\b} C^{\b\rho} (M^n)^\s_{~\rho} C_{\s\a}.
\ee
In the right hand side of both equations, the first term is
itself $SU(4)$ invariant. The remaining terms, which we denote as
$12X_1$ and $36X_2$, should combine into an $SU(4)$ invariant. To
see this, we introduce a new $SU(4)$ invariant term and decompose
it using the identities (\ref{id56}) and (\ref{fund6}), \be Z
&\equiv& \e_{\a\b\g\d} \e^{\a\r\s\t}
(M^{mn})^\b_{~\r}(M_m)^\g_{~\s}(M_n)^\d_{~\t} \; = \; 4X_2 -  X_3
+ 2X_4+2X_5 \,,
\\
&& X_3 \equiv (M_m)^\a_{~\b}(M_n)^\b_{~\g}C^{\g\rho}
(M^{mn})^\s_{~\rho} C_{\s\a} \,,
\\
&& X_4 \equiv (M_m)^\g_{~\g} (M^{nm})^\a_{~\b} C^{\b\rho}
(M_n)^\s_{~\rho} C_{\s\a} \,,
\\
&& X_5 \equiv (M_n)^\g_{~\g} (M^{nm})^\a_{~\b} C^{\b\rho}
(M_m)^\s_{~\rho} C_{\s\a}\,. \ee Inserting $f^{mn}{}_p t^p=\left[
t^m, t^n \right]$ into different $t^m$ factors in $X_1$, one can
show \be X_1 = X_2 + X_3 = -X_2 + X_4 = -X_2 + X_5, \ee from which
one can easily find $12X_1+36X_2=4Z$ as expected. Note also that
\begin{eqnarray}
Z &=&
     (M^{mn})^\beta_{~\rho}(M_m)^\gamma_{~\sigma}(M_n)^\delta_{~\tau}
 \left\{
  \delta^\rho  _\beta  \delta^\sigma_\gamma \delta^\tau  _\delta
  \pm (\mbox{5 other terms})
\right\}
 \nn\\ &=&
  4{\rm Tr}(M^{mn}M_nM_m) +2{\rm Tr}(M^{mn}M_mM_n)\,,
\end{eqnarray}
where the trace is with respect to the $SU(4)$ indices and the
fundamental identity was used. The potential term ${\cal L}_{\rm
pot}=-V=\pi^2 I/15$ finally becomes
\begin{eqnarray}
 I &=&
    -4f_{mnp}{\rm Tr}(M^mM^nM^p)
   +44{\rm Tr}(M_{mn}M^mM^n)
   +16{\rm Tr}(M^{mn}M_nM_m)
 \nn\\ &=&
       40{\rm Tr}(M^{mn}M_mM_n)+20{\rm Tr}(M^{mn}M_nM_m)
 \nn\\ &=&
       60{\rm Tr}(M^{mn}M_mM_n)-20f^{mnp}{\rm Tr}(M_mM_nM_p)\,, \nn\\
{\cal L}_{\rm pot}&=&
       4\pi^2{\rm Tr}(M^{mn}M_mM_n)
      -\frac{4\pi^2}3f^{mnp}{\rm Tr}(M_mM_nM_p)\,.
\end{eqnarray}

\section{Mass Deformation}

In this section, we present a supersymmetry preserving mass deformation
of the $\CN=5$ and $\CN=6$ theories.
It was shown in \cite{hl3p} that the extended Gaiotto-Witten theories
allow a mass deformation which preserves the whole $\CN=4$ supersymmetry
and $SO(4)$ $R$-symmetry.

The mass deformation adds the following terms to the Lagrangian \cite{hl3p},
\begin{eqnarray}
  {\cal L}_{\rm mass} &=&
 -\frac{\omega_{AB}}2\left(
  m^2\epsilon^{\alpha\beta}q_\alpha^A q_\beta^B
 +m^2\epsilon^{\dot\alpha\dot\beta}
  \tilde q_{\dot\alpha}^A\tilde q_{\dot\beta}^B
 +im\epsilon^{\dot\alpha\dot\beta}
  \psi_{\dot\alpha}^A\psi_{\dot\beta}^B
 -im\epsilon^{\alpha\beta}
  \tilde\psi_\alpha^A\tilde\psi_\beta^B
    \right)
 \nn\\ &&
 -\frac{2\pi}3mk_{mn}\left\{
    (\mu^m)_{\alpha\beta}(\mu^n)^{\beta\alpha}
   -(\tilde\mu^m)_{\dot\alpha\dot\beta}(\tilde\mu^n)^{\dot\beta\dot\alpha}
  \right\}\,,\label{mdef4}
\end{eqnarray}
and the supersymmetry transformation rules,
\begin{equation}
  \delta_{\rm mass}\psi^A_{\dot\alpha}~=~
  mq_\alpha^A\eta^\alpha_{~\dot\alpha},~~~~~
  \delta_{\rm mass}\tilde\psi_\alpha^A ~=~
  m\tilde q^A_{\dot\alpha}\eta_\alpha^{~\dot\alpha}.
\label{mdef4s}
\end{equation}
We will generalize this result to the $\CN=5$ and $\CN=6$
theories of this paper.
We will find that supersymmetries are all preserved, but the
$R$-symmetry gets partially broken.

\subsection*{$\CN=5$ mass deformation}

Using the notations introduced in Section \ref{n5},
the $\CN=4$ mass deformed Lagrangian (\ref{mdef4})
is rewritten as
\begin{eqnarray}
 {\cal L}_{\rm mass} &=&
 -\frac{\omega_{AB}}2\left(
   m^2\Phi_\alpha^AC^{\alpha\beta}\Phi_\beta^B
  -im\Psi_\alpha^A(C\Gamma_5)^{\alpha\beta}\Psi_\beta^B
  \right)
 \nn\\&&
 +\frac{2\pi m}3({\cal M}_m)_\alpha^{~\beta}({\cal M}^m)_\beta^{~\gamma}
  (\Gamma_5)_\gamma^{~\alpha}~,
\label{N5m}
\end{eqnarray}
where the matrix $\Gamma_5$ is defined by
$(\Gamma_5)_\alpha^{~\beta}={\rm diag}(+1,+1,-1,-1)$.
The mass deformation to the supersymmetry transformation rule is
\begin{equation}
 \delta_{\rm m}\Psi^A_\alpha =
 m(\Gamma_5)_\alpha^{~\gamma}\eta_\gamma^{~\beta}\Phi_\beta^A .
\label{N5ms}
\end{equation}
The explicit dependence on $\Gamma_5$ implies that the
$SO(5)$ $R$-symmetry is broken down to the $SO(4)$ subgroup.
Nevertheless, one can show that the mass deformed theory is invariant
under the whole $\CN=5$ supersymmetry deformed by (\ref{N5ms}).
This is consistent with previous results \cite{hll,x3} on mass
deformation of the $\CN=8$ BLG model of $SO(4)$ gauge group.
For the BLG model, the mass terms did not break any supersymmetry
despite the $R$-symmetry breaking.

\paragraph{Checking the supersymmetry}

Let us sketch the proof of the full $\CN=5$ invariance.
We work order by order in $m$.
The ${\cal O}(m^2)$ terms in $\delta{\cal L}$ arise from
$\delta$ of the boson mass term and $\delta_{\rm mass}$ of the
fermion mass term.
\begin{eqnarray}
 \delta{\cal L}|_{m^2} &=&
 -\omega_{AB}m^2\Phi_\alpha^AC^{\alpha\beta}\delta\Phi_\beta^B
 +im\omega_{AB}\Psi_\alpha^A(C\Gamma_5)^{\alpha\beta}
  \delta_{\rm mass}\Psi_\beta^B
 \nn\\ &=&
 -i\omega_{AB}m^2\Phi_\alpha^AC^{\alpha\beta}
  \eta_\beta^{~\gamma}\Psi_\gamma^B
 +im^2\omega_{AB}\Psi_\alpha^A(C\Gamma_5)^{\alpha\beta}
  (\Gamma_5)_\beta^{~\gamma}\eta_\gamma^{~\delta}\Phi_\delta^B
 ~=~ 0.
\end{eqnarray}
The ${\cal O}(m)$ terms fall into two types, one proportional
to $m\eta\Psi D\Phi$ and the other to $m\eta\Psi\Phi^3$.
The terms of the first type arise from $\delta$ of the fermionic
mass term and $\delta_{\rm mass}$ of the fermionic kinetic term,
and are easily shown to cancel each other.
The remaining ${\cal O}(m)$ terms are given by
\begin{eqnarray}
 \delta{\cal L}|_m
 &=&
 ~~\frac{2\pi im}3
 \left(
   (\Gamma_5)_\alpha^{~\beta}\eta_\beta^{~\gamma}
   ({\cal M}_m)_\gamma^{~\delta}({\cal J}^m)_\delta^{~\alpha}
 -2(\Gamma_5)_\alpha^{~\beta}({\cal M}_m)_\beta^{~\gamma}
   \eta_\gamma^{~\delta}({\cal J}^m)_\delta^{~\alpha}
 \right)
 \nn\\&&
+\frac{2\pi im}3
 \left(
  2(\Gamma_5)_\gamma^{~\alpha}({\cal M}_m)_\alpha^{~\beta}
  \eta_\beta^{~\delta}({\cal J}^m)_\delta^{~\gamma}
 +2(\Gamma_5)_\gamma^{~\alpha}({\cal M}_m)_\alpha^{~\beta}
  ({\cal J}^m)_\beta^{~\delta}\eta_\delta^{~\gamma}
 \right)
 \nn\\&&
+\frac{2\pi im}3
 \Big(
 3(\Gamma_5)_\gamma^{~\epsilon}\eta_\epsilon^{~\rho}
  ({\cal M}_m)_\rho^{~\alpha}({\cal J}^m)_\alpha^{~\gamma}
 -6(\Gamma_5)_\epsilon^{~\gamma}({\cal J}^m)_\gamma^{~\alpha}
   ({\cal M}_m)_\alpha^{~\rho}\eta_\rho^{~\epsilon}
 \Big)~,
\end{eqnarray}
where the three lines in the right hand side correspond respectively to
$\delta$ of the fermion mass term,
$\delta$ of the quartic potential term and
$\delta_{\rm mass}$ of the Yukawa interaction.
It appears difficult to show that this vanishes, but since
we know it vanishes when $\eta_\alpha^{~\beta}$ is proportional
to $\Gamma_1,\cdots,\Gamma_4$, we only need to check that it
vanishes when $\eta\sim\Gamma_5$.

\subsection*{$\CN=6$ mass deformation}

Following the lifting procedure in section \ref{n6}, one can show that
the mass deformed Lagrangian is given by
\begin{eqnarray}
 {\cal L}_{\rm mass} &=&
 -m^2\bar\Phi_A^\alpha\Phi^A_\alpha
 +im\bar\Psi_{\alpha A}(C\Gamma_5C)^\alpha_{~\beta}\Psi^{A\beta}
 \nn\\&&
 +\frac{2\pi m}3(M^m)^\alpha_{~\beta}(M_m)^\gamma_{~\delta}
  \Big\{-2\delta_\alpha^{~\beta}(\Gamma_5)_\gamma^{~\delta}
        + (\Gamma_5C)_{\alpha\gamma}C^{\beta\delta}
        +  C_{\alpha\gamma}(C\Gamma_5)^{\beta\delta}
     \Big\}~.
\end{eqnarray}
We regard $\Phi_\alpha^A$ as a $SO(6)$ left-handed spinor and
$\Psi^{A\alpha}$ as a right-handed spinor.
Introducing the six-dimensional Gamma matrices
\begin{equation}
 \hat\Gamma_I=\left(\begin{array}{cc}
 0&(\bar\rho_I)_{\alpha\beta}\\(\rho_I)^{\alpha\beta}&0\end{array}\right),
 ~~~~~
 \rho_I=(C\Gamma_I,iC),~~~~
 \bar\rho_I=(\Gamma_IC,-iC)\,,
\end{equation}
one can write ${\cal L}_{\rm mass}$ in the following form
\begin{eqnarray}
 {\cal L}_{\rm m} &=&
 -m^2\bar\Phi_A^\alpha\Phi^A_\alpha
 -m\bar\Psi_{\alpha A}(\rho_{56})^\alpha_{~\beta}\Psi^{A\beta}
 \nn\\&&
 +\frac{2\pi im}3(M^m)^\alpha_{~\beta}(M_m)^\gamma_{~\delta}
  \Big\{2\delta_\alpha^{~\beta}(\bar\rho_{56})_\gamma^{~\delta}
        -(\bar\rho_5)_{\alpha\gamma}(\rho_6)^{\beta\delta}
        +(\bar\rho_6)_{\alpha\gamma}(\rho_5)^{\beta\delta}
     \Big\}\,.
\end{eqnarray}
The deformation to the supersymmetry transformation rule is
\begin{equation}
 \delta_{\rm mass}\Psi^{\alpha A}
 ~=~ m(C\Gamma_5 C)^\alpha_{~\beta}\eta^{\beta\gamma}\Phi_\gamma^A
 ~=~ im(\rho_{56})^\alpha_{~\beta}\eta^{\beta\gamma}\Phi_\gamma^A~.
\end{equation}
The mass deformation breaks the $SO(6)$ $R$-symmetry
down to $SO(4)\times SO(2)$.
But this deformation preserves all $\CN=6$ supersymmetry, since the
$SO(2)$ relates the sixth supersymmetry with the fifth one which has
been shown to be the symmetry.
See \cite{bena,ABJM} for related discussions.


\vskip 2cm





\begin{thebibliography}{99}


\bibitem{sch}
  J.~H.~Schwarz,
  ``Superconformal Chern-Simons theories,''
  JHEP {\bf 0411} (2004) 078
  [arXiv:hep-th/0411077].

\bibitem{BL1}
  J.~Bagger and N.~Lambert,
  ``Modeling multiple M2's,''
  Phys.\ Rev.\  D {\bf 75} (2007) 045020
  [arXiv:hep-th/0611108].


 \bibitem{BL2}
  J.~Bagger and N.~Lambert,
  ``Gauge Symmetry and Supersymmetry of Multiple M2-Branes,''
  Phys.\ Rev.\  D {\bf 77} (2008) 065008
  [arXiv:0711.0955 [hep-th]].

\bibitem{BL3}
  J.~Bagger and N.~Lambert,
  ``Comments On Multiple M2-branes,''
  JHEP {\bf 0802} (2008) 105
  [arXiv:0712.3738 [hep-th]].



\bibitem{gus1}
  A.~Gustavsson,
  ``Algebraic structures on parallel M2-branes,''
  arXiv:0709.1260 [hep-th].

\bibitem{gus2}
  A.~Gustavsson,
  ``Selfdual strings and loop space Nahm equations,''
  arXiv:0802.3456 [hep-th].



 \bibitem{sc2}
   M.~A.~Bandres, A.~E.~Lipstein and J.~H.~Schwarz,
   ``N = 8 Superconformal Chern--Simons Theories,''
  arXiv:0803.3242 [hep-th].

\bibitem{rams}
  M.~Van Raamsdonk,
  ``Comments on the Bagger-Lambert theory and multiple M2-branes,''
  arXiv:0803.3803 [hep-th].


\bibitem{tong}
  N.~Lambert and D.~Tong,
  ``Membranes on an Orbifold,''
  arXiv:0804.1114 [hep-th].
 


\bibitem{mukhi}
  J.~Distler, S.~Mukhi, C.~Papageorgakis and M.~Van Raamsdonk,
  ``M2-branes on M-folds,''
  arXiv:0804.1256 [hep-th].



\bibitem{ABJM}
  O.~Aharony, O.~Bergman, D.~L.~Jafferis and J.~Maldacena,
  ``N=6 superconformal Chern-Simons-matter theories, M2-branes and their
  gravity duals,''
  arXiv:0806.1218 [hep-th].

\bibitem{gw}
  D.~Gaiotto and E.~Witten,
  ``Janus Configurations, Chern-Simons Couplings, And The Theta-Angle in N=4
  Super Yang-Mills Theory,''
  arXiv:0804.2907 [hep-th].


\bibitem{hl3p}
  K.~Hosomichi, K.~M.~Lee, S.~Lee, S.~Lee and J.~Park,
  ``N=4 Superconformal Chern-Simons Theories with Hyper and Twisted Hyper
  Multiplets,''
  arXiv:0805.3662 [hep-th].



 \bibitem{gy}
   D.~Gaiotto and X.~Yin,
  ``Notes on superconformal Chern-Simons-matter theories,''
  JHEP {\bf 0708} (2007) 056
   [arXiv:0704.3740 [hep-th]].

\bibitem{a0}
  C.~Ahn, K.~Oh and R.~Tatar,
  ``Branes, orbifolds and the three dimensional N = 2 SCFT in the large N
  limit,''
  JHEP {\bf 9811} (1998) 024
  [arXiv:hep-th/9806041].

\bibitem{hana}
  A.~Hanany, N.~Mekareeya and A.~Zaffaroni,
  ``Partition Functions for Membrane Theories,''
  arXiv:0806.4212 [hep-th].





 \bibitem{x3}
   J.~Gomis, A.~J.~Salim and F.~Passerini,
   ``Matrix Theory of Type IIB Plane Wave from Membranes,''
   arXiv:0804.2186 [hep-th].


\bibitem{hll}
    K.~Hosomichi, K.~M.~Lee and S.~Lee,
   ``Mass-Deformed Bagger-Lambert Theory and its BPS Objects,''
   arXiv:0804.2519 [hep-th].
 



\bibitem{kek}
  Y.~Honma, S.~Iso, Y.~Sumitomo and S.~Zhang,
  ``Scaling limit of N=6 superconformal Chern-Simons theories and Lorentzian
  Bagger-Lambert theories,''
  arXiv:0806.3498 [hep-th].
 




\bibitem{bf1}
  J.~Gomis, G.~Milanesi and J.~G.~Russo,
  ``Bagger-Lambert Theory for General Lie Algebras,''
  arXiv:0805.1012 [hep-th].
 

\bibitem{bf2}
  S.~Benvenuti, D.~Rodriguez-Gomez, E.~Tonni and H.~Verlinde,
  ``N=8 superconformal gauge theories and M2 branes,''
  arXiv:0805.1087 [hep-th].


\bibitem{bf3}
  P.~M.~Ho, Y.~Imamura and Y.~Matsuo,
  ``M2 to D2 revisited,''
  arXiv:0805.1202 [hep-th].



\bibitem{bx1}
  M.~A.~Bandres, A.~E.~Lipstein and J.~H.~Schwarz,
  ``Ghost-Free Superconformal Action for Multiple M2-Branes,''
  arXiv:0806.0054 [hep-th].

\bibitem{bb1}
  J.~Gomis, D.~Rodriguez-Gomez, M.~Van Raamsdonk and H.~Verlinde,
  ``The Superconformal Gauge Theory on M2-Branes,''
  arXiv:0806.0738 [hep-th].

\bibitem{bx2}
  B.~Ezhuthachan, S.~Mukhi and C.~Papageorgakis,
  ``D2 to D2,''
  arXiv:0806.1639 [hep-th].


\bibitem{bb2}
  S.~Cecotti and A.~Sen,
  ``Coulomb Branch of the Lorentzian Three Algebra Theory,''
  arXiv:0806.1990 [hep-th].





\bibitem{a2}
  M.~Benna, I.~Klebanov, T.~Klose and M.~Smedback,
  ``Superconformal Chern-Simons Theories and AdS$_4$/CFT$_3$ Correspondence,''
  arXiv:0806.1519 [hep-th].



\bibitem{ima}
  Y.~Imamura and K.~Kimura,
  ``Coulomb branch of generalized ABJM models,''
  arXiv:0806.3727 [hep-th].


\bibitem{a8}
  J.~Bhattacharya and S.~Minwalla,
  ``Superconformal Indices for ${\cal N}=6$ Chern Simons Theories,''
 arXiv:0806.3251 [hep-th].


\bibitem{a10}
  T.~Nishioka and T.~Takayanagi,
  ``On Type IIA Penrose Limit and N=6 Chern-Simons Theories,''
  arXiv:0806.3391 [hep-th].

  


\bibitem{mz}
  J.~A.~Minahan and K.~Zarembo,
  ``The Bethe ansatz for superconformal Chern-Simons,''
  arXiv:0806.3951 [hep-th].



\bibitem{adi}
  A.~Armoni and A.~Naqvi,
  ``A Non-Supersymmetric Large-N 3D CFT And Its Gravity Dual,''
  arXiv:0806.4068 [hep-th].



\bibitem{lyee}
  K.~M.~Lee and H.~U.~Yee,
  ``New AdS(4) x X(7) geometries with N = 6 in M theory,''
  JHEP {\bf 0703} (2007) 012
  [arXiv:hep-th/0605214].


\bibitem{mc1}
  S.~Lee,
  ``Superconformal field theories from crystal lattices,''
  Phys.\ Rev.\  D {\bf 75} (2007) 101901
  [arXiv:hep-th/0610204].

\bibitem{mc2}
 S.~Lee, S.~Lee and J.~Park,
  ``Toric AdS$_4$/CFT$_3$ duals and M-theory crystals,''
  JHEP {\bf 0705} (2007) 004
  [arXiv:hep-th/0702120].

\bibitem{mc3}
  S.~Kim, S.~Lee, S.~Lee and J.~Park,
 ``M2-brane Probe Dynamics and Toric Duality,''
  Nucl.\ Phys.\  B {\bf 797} (2008) 340
  [arXiv:0705.3540 [hep-th]].


\bibitem{a1}
  C.~Ahn,
  ``Holographic Supergravity Dual to Three Dimensional N=2 Gauge Theory,''
  arXiv:0806.1420 [hep-th].




\bibitem{ggy}
  D.~Gaiotto, S.~Giombi and X.~Yin,
  ``Spin Chains in N=6 Superconformal Chern-Simons-Matter Theory,''
  arXiv:0806.4589 [hep-th].


\bibitem{Elitzur:1998ju}
  S.~Elitzur, A.~Giveon, D.~Kutasov and D.~Tsabar,
  ``Branes, orientifolds and chiral gauge theories,''
  Nucl.\ Phys.\  B {\bf 524}, 251 (1998)
  [arXiv:hep-th/9801020].



\bibitem{kw}
  I.~R.~Klebanov and E.~Witten,
  ``Superconformal field theory on threebranes at a Calabi-Yau  singularity,''
  Nucl.\ Phys.\  B {\bf 536} (1998) 199
  [arXiv:hep-th/9807080].

\bibitem{dimerori}
 S. Franco , A. Hanany , D. Krefl , J. Park , A. M. Uranga and D. Vegh,
 ``Dimers and orientifolds,''
 JHEP {\bf 0709} (2007) 075
 [arXiv:0707.0298 [hep-th]].


\bibitem{polchinskiz2}
 M. Berkooz, R. Leigh, J. Polchinski , J. H. Schwarz , N. Seiberg and E. Witten,
 ``Anomalies, dualities, and topology of D = 6 N=1 superstring vacua,''
  Nucl.\ Phys.\ B {\bf 475} (1996) 115
  [arXiv: hep-th/9605184].


\bibitem{bena}
  I.~Bena,
  ``The M-theory dual of a 3 dimensional theory with reduced supersymmetry,''
  Phys.\ Rev.\  D {\bf 62} (2000) 126006
  [arXiv:hep-th/0004142].

\end{thebibliography}
\end{document}